\documentclass[12pt,notitlepage]{iopart}
\usepackage[dvips]{color}
\usepackage{iopams,amsfonts,amssymb,latexsym,epsf,bm} 
\expandafter\let\csname equation*\endcsname\relax
\expandafter\let\csname endequation*\endcsname\relax
\usepackage{amsmath}
\usepackage{graphicx}

\newcommand{\ket}[1]{|#1 \rangle}

\newcommand{\av}{{\mathbf{a}}}
\newcommand{\bv}{{\mathbf{b}}}
\newcommand{\ev}{{\mathbf{e}}}
\newcommand{\rv}{{\mathbf{r}}}

\newcommand{\kv}{{\mathbf{k}}}
\newcommand{\pv}{{\mathbf{p}}}
\newcommand{\qv}{{\mathbf{q}}}
\newcommand{\mv}{{\mathbf{m}}}
\newcommand{\nv}{{\mathbf{n}}}
\newcommand{\sv}{\mathbf{s}}
\newcommand{\uv}{{\mathbf{u}}}

\newcommand{\wv}{{\mathbf{w}}}
\newcommand{\xv}{{\mathbf{x}}}
\newcommand{\Xv}{{\mathbf{X}}}

\newcommand{\zerov}{{\mathbf{0}}}

\newcommand{\Av}{{\mathbf{A}}}
\newcommand{\Bv}{{\mathbf{B}}}

\newcommand{\Kv}{{\mathbf{K}}}
\newcommand{\Gv}{{\mathbf{G}}}
\newcommand{\kvt}{\tilde{\mathbf{k}}}
\newcommand{\Sv}{{\mathbf{S}}}
\newcommand{\Pv}{{\mathbf{P}}}

\newcommand{\psih}{\hat{\psi}}

\newcommand{\taub}{\bar{\tau}}

\newcommand{\bt}{\tilde{b}}
\newcommand{\Vt}{\tilde{V}}

\newcommand{\Ecal}{\mathcal{E}}

\newcommand{\Ucal}{\mathcal{U}}
\newcommand{\Vcal}{\mathcal{V}}

\newcommand{\pdif}[1]{ \frac{\partial}{\partial #1} }

\newcommand{\pdiff}[2]{ \frac{\partial #1}{\partial #2} }

\newcommand{\Zbb}{\mathbb{Z}}

\newcommand{\ua}{\uparrow}
\newcommand{\da}{\downarrow}

\newcommand{\Nvor}{N_\mathrm{v}}

\newcommand{\mix}{\mathrm{mix}}

\newcommand{\thetas}{\theta_\mathrm{sing}}
\newcommand{\jvor}{j_\mathrm{vor}}
\newcommand{\sing}{\mathrm{sing}}
\newcommand{\reg}{\mathrm{reg}}
\newcommand{\eff}{\mathrm{eff}}
\newcommand{\elastic}{\mathrm{el}}
\newcommand{\ac}{\mathrm{ac}}
\newcommand{\op}{\mathrm{op}}

\newcommand{\Tapara}{T^\mathrm{(AP)}}

\begin{document}
\title[]{
Collective modes of vortex lattices\\ in two-component Bose-Einstein condensates\\ under synthetic gauge fields
}
\author{Takumi Yoshino$^1$, Shunsuke Furukawa$^1$, Sho Higashikawa$^1$,\\ and Masahito Ueda$^{1,2}$}
\address{
$^1$Department of Physics, University of Tokyo, 7-3-1 Hongo, Bunkyo-ku, Tokyo 113-0033, Japan\\
$^2$RIKEN Center for Emergent Matter Science (CEMS), Wako, Saitama 351-0198, Japan
}
\ead{yoshino@cat.phys.s.u-tokyo.ac.jp and furukawa@cat.phys.s.u-tokyo.ac.jp}


\vspace{1pc}

\begin{abstract}
We study collective modes of vortex lattices in two-component Bose-Einstein condensates 
subject to synthetic magnetic fields in mutually parallel or antiparallel directions. 
By means of the Bogoliubov theory with the lowest-Landau-level approximation, we numerically calculate the excitation spectra for a rich variety of vortex lattices 
that appear commonly for parallel and antiparallel synthetic fields. 
We find that in all of these cases, there appear two distinct modes with linear and quadratic dispersion relations at low energies, 
which exhibit anisotropy reflecting the symmetry of each lattice structure. 
Remarkably, the low-energy spectra for the two types of fields are found to be related to each other by simple rescaling 
when vortices in different components overlap owing to an intercomponent attraction. 
These results are consistent with an effective field theory analysis. 
However, the rescaling relations break down for interlaced vortex lattices appearing with an intercomponent repulsion, 
indicating a nontrivial effect of an intercomponent vortex displacement beyond the effective field theory. 
We also find that high-energy parts of the excitation bands exhibit line or point nodes 
as a consequence of a fractional translation symmetry present in some of the lattice structures.
\end{abstract}


\section{Introduction}

Formation of quantized vortices under rotation is a hallmark of superfluidity. 
When quantized vortices proliferate under rapid rotation, they organize into a regular lattice owing to their mutual repulsion. 
The resulting triangular vortex lattice structure was originally predicted by Abrikosov \cite{Abrikosov56} for type-II superconductors in a magnetic field, 
and observed in superconducting materials \cite{Essmann67}, superfluid $^4$He \cite{Yarmchuk82,Donnelly91}, 
and Bose-Einstein condensates (BEC) \cite{AboShaeer01,Engels02,Schweikhard04_1comp} and fermionic superfluids \cite{Zwierlein05} of ultracold atoms. 
In ultracold atomic gases, in particular, the rotation frequency can be tuned over a wide range, 
and the equilibrium and dynamical properties of vortex lattices can be investigated in considerable detail \cite{Stock05,Cooper08,Fetter09}. 
Rotation can be viewed as the standard way to induce a {\it synthetic} gauge field for neutral atoms 
since the Hamiltonian in the rotating frame of reference is equivalent to that of charged particles in a uniform magnetic field. 
Notably, experimental techniques for producing synthetic gauge fields via optical dressing of atoms  
have also been developed over the past decade \cite{Dalibard11, Goldman14}, 
and a successful application of these techniques led to the creation of around $10$ vortices in a BEC without rotating the gas \cite{Lin09}. 

Throughout this paper, we assume that a BEC is confined in a three-dimensional harmonic potential 
and that the interparticle interaction is so strong that the BEC at rest is in the Thomas-Fermi regime. 
A BEC under rotation (or in a synthetic magnetic field) undergoes different regimes with increasing the rotation frequency $\Omega$ \cite{Fetter09}. 
When a BEC rotates slowly, the size of the vortex core is much smaller than the intervortex separation. 
In this regime, the spatial variation of the BEC density, $\nabla |\Psi|$, can be ignored, 
and the Thomas-Fermi approximation is still applicable \cite{Butts99}. 
This regime is called the mean-field Thomas-Fermi regime. 
With increasing $\Omega$, the intervortex separation decreases and eventually becomes comparable with the size of a vortex core.  
Then the BEC flattens to an effectively two-dimensional (2D) system, and the interaction energy per particle becomes small compared with the kinetic energy per particle. 
It is thus reasonable to assume that atoms reside in the lowest-Landau-level (LLL) manifold for the motion in the 2D plane 
and to perform the mean-field calculation in this manifold \cite{Ho01,Baym05}. 
This regime is called the mean-field LLL regime \cite{Cooper08}. 
As $\Omega$ is further increased, the mean-field description breaks down, and the system is expected to enter a highly correlated regime. 
In particular, in a regime where the number of vortices $\Nvor$ becomes comparable with the number of atoms $N$, 
it has been predicted that the vortex lattice melts and a variety of quantum Hall states appear 
at integral and fractional values of the filling factor $\nu:=N/\Nvor$ \cite{Cooper08, Wilkin98, Cooper01}.



A vortex lattice supports an elliptically polarized oscillatory mode, 
which was predicted by Tkachenko \cite{Tkachenko66a,Tkachenko66b,Tkachenko69} 
and observed in superfluid $^4$He \cite{Andereck82}. 
While Tkachenko's original work predicted a linear dispersion relation for an incompressible fluid, 
a number of theoretical studies have been done to take into account a finite compressibility of the fluid 
\cite{Sonin76, Williams77, Baym83_1,Chandler86_2,Sonin87}. 
It has been shown that the compressibility leads to hybridization with sound waves 
and qualitatively changes the dispersion relation into a quadratic form for small wave vectors. 
Collective modes of a vortex lattice have been observed over a wide range of rotation frequencies in a harmonically trapped BEC \cite{Coddington03}. 
Theoretical analyses of the observed modes have been conducted
with the hydrodynamic theory \cite{Baym03, Cozzini04, Sonin05_cont} and the Gross-Pitaevskii (GP) mean-field theory \cite{Mizushima04, Baksmaty04}. 
For a uniform BEC in the mean-field LLL regime, the dispersion relation of the Tkachenko mode can analytically be obtained within the Bogoliubov theory, 
and it is found to take a quadratic form \cite{Sinova02, Matveenko11, Kwasigroch12}. 
Effective field theory for the Tkachenko mode has been developed in Refs.\ \cite{WatanabeMurayama13,Moroz18}. 




\begin{figure}
 \centering
    \includegraphics[width=16cm , angle=0]{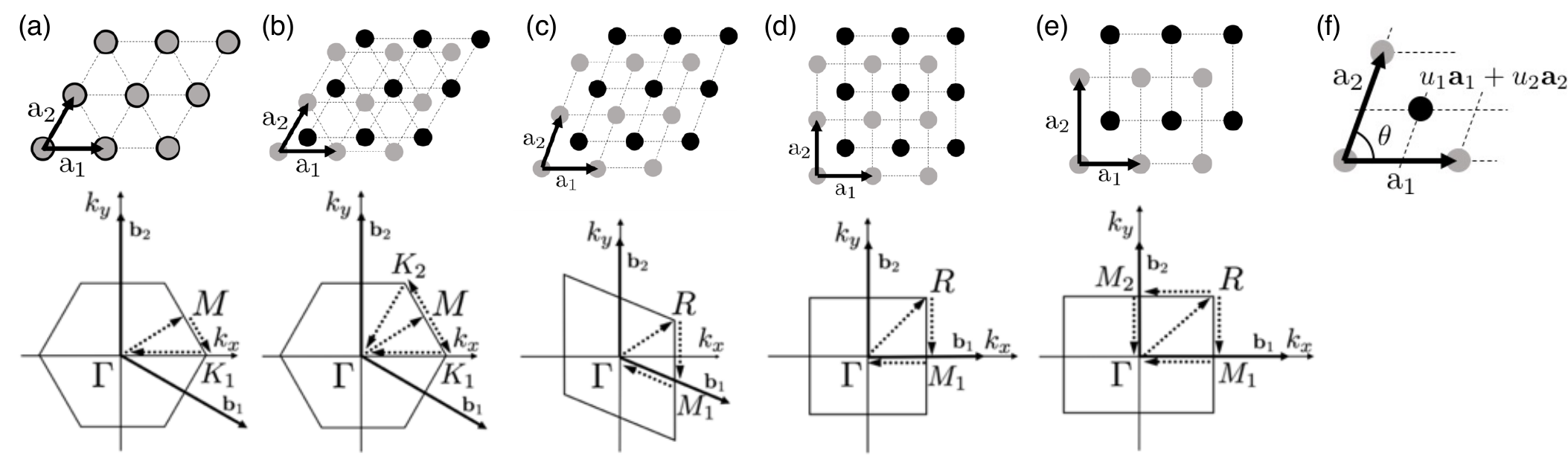}
 \caption{\label{fig:vorlat}
Upper panels: vortex-lattice structures in two-component BECs in synthetic magnetic fields \cite{Mueller02,Kasamatsu03,Kasamatsu05,Oktel06}. 
Within the GP mean-field theory, the same phase diagrams are obtained for both the parallel- and antiparallel-field cases \cite{Furukawa14}. 
Five different structures appear as the ratio of the coupling constants, $g_{\ua\da}/g$, is varied: 
(a) overlapping triangular lattices ($-1<g_{\ua\da}/g<0$), (b) interlaced triangular lattices ($0<g_{\ua\da}/g<0.1724$), 
(c) rhombic lattices ($0.1724 < g_{\ua\da}/g<0.3733$), (d) square lattices ($0.3733 < g_{\ua\da}/g<0.9256$), and (e) rectangular lattices ($0.9256 < g_{\ua\da}/g<1$). 
Here, $g_{\ua\da}$ is the intercomponent coupling constant, and $g$ is the intracomponent one which is assumed to be the same for both components.  
Black (grey) circles indicate the vortex positions in the spin-$\ua$ ($\da$) component. 
As shown in (f), each lattice structure is characterized by the primitive vectors $\av_1=(a,0)$ and $\av_2=b(\cos\theta,\sin\theta)$ 
satisfying $ab\sin\theta=2\pi\ell^2$ [see Eq.~\eqref{eq:quan_flux}], 
and the vortex displacement $u_1\av_1+u_2\av_2$ of one component relative to the other. 
The angle $\theta$ (the aspect ratio $b/a$) varies continuously in the rhombic-lattice (rectangular-lattice) phase, as shown in Refs.\ \cite{Mueller02,Oktel06}. 
Lower panels: the first Brillouin zone corresponding to each lattice structure placed above. 
The reciprocal primitive vectors are given by $\bv_1=(b\sin\theta,-b\cos\theta)/\ell^2$ and $\bv_2 = (0,a)/\ell^2$ [see Eq.\ \eqref{eq:bv12}]. 
Uppercase letters indicate high-symmetry points. 
Excitation spectra presented in Fig.~\ref{fig:spectrum} are calculated along the paths indicated by dotted arrows.
}
\end{figure}

The properties of vortex latices can further be enriched in {\it multicomponent} BECs, 
such as those made up of different hyperfine spin states of identical atoms. 
For two-component BECs under rotation, GP mean-field calculations have shown that 
several different types of vortex lattices appear 
as the ratio of the intercomponent coupling $g_{\ua\da}$ to the intracomponent one $g>0$ is varied (see Fig.\ \ref{fig:vorlat})
\cite{Mueller02,Kasamatsu03,Kasamatsu05}. 
Among them, interlaced square vortex lattices  [Fig.\ \ref{fig:vorlat}(d)], which are unique to these systems, 
have been observed experimentally \cite{Schweikhard04_2comp}. 
Furthermore, optical dressing techniques can produce a variety of (possibly non-Abelian) gauge fields in multicomponent gases \cite{Dalibard11, Goldman14, Lin11,Zhai12_review}. 
In particular, mutually antiparallel synthetic magnetic fields have been induced in two-component BECs, leading to the observation of the spin Hall effect \cite{Beeler13}. 
If the antiparallel fields are made even higher, such systems are expected to show a rich phase diagram 
consisting of vortex lattices and (fractional) quantum spin Hall states \cite{Liu07,Fialko14,Furukawa14}. 
Notably, it has been shown {\it within} the GP mean-field theory that BECs in antiparallel magnetic fields exhibit the {\it same} vortex-lattice phase diagram 
as BECs in parallel magnetic fields \cite{Furukawa14} (see also Sec.\ \ref{sec:systems}). 
It is thus interesting to ask whether and how the difference between the two types of systems arises in other properties such as collective modes. 
In this context, it is worth noting that in the quantum Hall regime, which is far beyond the mean-field description, 
the two types of systems exhibit markedly different phase diagrams \cite{Furukawa14,Furukawa13,Regnault13,Geraedts17,Furukawa17}, 
which has been interpreted in light of pseudopotentials and entanglement formation \cite{Furukawa17}.

In this paper, we study collective modes of vortex lattices in two-component BECs 
in parallel and antiparallel synthetic magnetic fields in the mean-field LLL regime. 
On the basis of the Bogoliubov theory with the LLL approximation, we numerically calculate excitation spectra for all the vortex-lattice structures shown in Fig.\ \ref{fig:vorlat}. 
We find that in all the cases, there appear two distinct modes with quadratic and linear dispersion relations at low energies, 
which originate from in-phase and anti-phase (i.e., $\pi$-phase difference) oscillations of vortices of the two components, respectively.  
The obtained dispersion relations show anisotropy reflecting the symmetry of each lattice structure. 
Remarkably, the low-energy spectra for the two types of synthetic fields are related to each other by simple rescaling 
in the case of overlapping vortex lattices [Fig.\ \ref{fig:vorlat}(a)] that appear for an intercomponent attraction. 
These results are consistent with an effective field theory analysis for low energies, 
which is a generalization of Ref.~\cite{WatanabeMurayama13} aided with symmetry consideration of the elastic energy of a vortex lattice. 
However, the rescaling relations are found to break down for interlaced vortex lattices [Fig.\ \ref{fig:vorlat}(b)-(e)] that appear for an intercomponent repulsion, 
presumably due to a nontrivial effect of a vortex displacement between the components beyond the effective field theory. 
We also find some interesting features of the excitation bands at high energies, such as line and point nodes, 
which arise from ``fractional'' translation symmetries or special structures of the Bogoliubov Hamiltonian matrix. 

Here we comment on some related studies. 
Ke\c{c}eli and Oktel \cite{Oktel06} have studied collective excitation spectra in two-component BECs in parallel fields by means of the hydrodynamic theory, 
and predicted the appearance of two low-energy modes with linear and quadratic dispersion relations similar to ours. 
Our calculation is based on the Bogoliubov theory, provides unbiased results for weak interactions,  
and also contains information on the higher-energy part of the spectra. 
Furthermore, in the effective field theory analysis, 
we point out a term missing in Ref.\ \cite{Oktel06}, which is responsible for 
the anisotropy of the quadratic dispersion relation for interlaced triangular lattices [Fig.\ \ref{fig:vorlat}(b)]. 
We also note that Woo {\it et al.} \cite{Woo07} have numerically investigated 
excitation spectra in rotating two-component BECs in a harmonic trap, 
and have identified a variety of excitations such as Tkachenko modes and surface waves. 

The rest of this paper is organized as follows. 
In Sec.\ \ref{sec:Bogo}, we introduce the systems that we study in this paper, 
and formulate the problem in terms of the Bogoliubov theory in the LLL basis. 
We then present our numerical results of Bogoliubov excitation spectra. 
In Sec.\ \ref{sec:Effect}, we use an effective field theory to derive analytical formulae of low-energy excitation spectra. 
In particular, we find remarkable rescaling relations between the spectra for the two types of synthetic magnetic fields. 
In Sec.\ \ref{sec:aniso}, we analyze the anisotropy of low-energy excitation spectra using the numerical data, 
and discuss its consistency with the effective field theory. 
In Sec.\ \ref{sec:summary}, we summarize the main results and discuss the outlook for future studies. 
In \ref{app:mag_Bloch_theta}, we derive expressions of the LLL magnetic Bloch states 
(the basis states used throughout this paper) in terms of Jacobi's theta functions; 
such expressions are used when plotting density profiles of excitation modes in Sec.~\ref{sec:Bogo} and \ref{app: element}. 
In \ref{app: calc}, we describe the derivation of the matrix elements of the interaction used in Sec.\ \ref{sec:Bogo}. 
In \ref{app:fractrans}, we give precise definitions of the fractional translation operators used in Sec.\ \ref{sec:Bogo}. 
In \ref{app: element}, we discuss some features of the Bogoliubov excitation spectra at high-symmetry points (found in Sec.\ \ref{sec:Bogo})
by using the data of the Bogoliubov Hamiltonian matrix and the density profiles of the excitation modes. 
In \ref{app:elastic}, we present symmetry consideration of the elastic energy of vortex lattices, which is used in Sec.\ \ref{sec:Effect}.

\section{Bogoliubov analysis of excitation spectra}\label{sec:Bogo}

In this section, we introduce the systems that we study in this paper, 
and formulate the problem in terms of the Bogoliubov theory with the LLL approximation.
Our formulation is closely related to those in Refs.\ \cite{Sinova02, Matveenko11, Kwasigroch12}. 
In particular, the LLL magnetic Bloch states \cite{Kwasigroch12,Rashba97,Burkov10}, which have a periodic pattern of zeros, play a crucial role here. 
We then present our numerical results of Bogoliubov excitation spectra and discuss their low- and high-energy characteristics. 

\subsection{Systems}\label{sec:systems}

We consider a system of a 2D pseudospin-$\frac12$ Bose gas having two hyperfine spin states (labeled by $\alpha=\ua,\da$). 
The spin-$\alpha$ component is subject to a synthetic magnetic field $B_\alpha$ in the $z$ direction. 
In the case of a gas rotating with an angular frequency $\Omega$, 
parallel fields $B_\ua=B_\da= 2M\Omega/q$ are induced in the two components in the rotating frame of reference, 
where $M$ and $q$ are the mass and the fictitious charge, respectively, of a neutral atom. 
An optical dressing technique of Ref.~\cite{Beeler13}, in contrast, can be used to produce antiparallel fields $B_\ua= -B_\da$. 
We focus on a central region of the system where the atomic density is sufficiently uniform and the effect of the harmonic potential can be ignored. 
In the second-quantized form, the Hamiltonian of the system is given by 
\begin{equation}\label{eq:2compHami}
\begin{split}
 H &= H_\mathrm{kin} + H_\mathrm{int}\\
 &=\sum_{\alpha=\ua,\da}  \int d^2\rv~ \psih_{\alpha}^{\dag}(\rv) \frac{(\pv-q\Av_\alpha)^2}{2M} \psih_\alpha(\rv)
 +\sum_{\alpha,\beta} \frac{g_{\alpha\beta}}{2} \int d^2\rv~ \psih^\dag_\alpha(\rv)\psih^\dag_\beta(\rv)\psih_\beta(\rv)\psih_\alpha(\rv),
\end{split}
\end{equation}
where $\rv=(x,y)$ is the coordinate on the 2D plane, $\pv=-i\hbar(\partial_x,\partial_y)$ is the momentum, 
and $\psih_\alpha(\rv)$ is the bosonic field operator for the spin-$\alpha$ component satisfying the commutation relations 
$[\psih_\alpha(\rv),\psih_\beta^\dag(\rv')]=\delta_{\alpha\beta}\delta^{(2)}(\rv- \rv')$ and $[\psih_\alpha(\rv),\psih_\beta(\rv')]= [\psih_\alpha^\dag(\rv),\psih_\beta^\dag(\rv')]=0$. 
The gauge field for the spin-$\alpha$ component is given by 
\begin{equation}
 \Av_\alpha=\frac{B_\alpha}{2} \ev_z\times\rv = \epsilon_\alpha \frac{B}{2}(-y,x), 
\end{equation}
where we assume $B>0$ 
and $\epsilon_\ua=\epsilon_\da=1$ ($\epsilon_\ua=-\epsilon_\da=1$) for parallel (antiparallel) fields. 
For a 2D system of area $A$, the number of magnetic flux quanta piercing each component (or the number of vortices) is given by $\Nvor =A/(2\pi\ell^2)$, 
where $\ell=\sqrt{\hbar/qB}$ is the magnetic length. 
The total number of atoms is given by $N=N_\ua+N_\da$, where $N_\alpha$ is the number of spin-$\alpha$ bosons.

In the Hamiltonian \eqref{eq:2compHami}, we assume a contact interaction between atoms. 
For a gas tightly confined in a harmonic potential with frequency $\omega_z$ in the $z$ direction, 
the effective coupling constants in the 2D plane are given by 
$g_{\alpha\alpha}=a_\alpha\sqrt{8\pi \hbar^3\omega_z/M}$ and 
$g_{\ua\da}=g_{\da\ua}=a_{\ua\da}\sqrt{8\pi \hbar^3\omega_z/M}$,\footnote{
These are obtained by multiplying the coupling constants 
$g_\alpha^\mathrm{(3D)}=4\pi\hbar^2 a_\alpha/M$ and $g_{\ua\da}^\mathrm{(3D)}=4\pi\hbar^2 a_{\ua\da}/M$ for the 3D contact interactions 
by the factor $\sqrt{M\omega_z/(2\pi \hbar)}$. 
This factor arises from the restriction to the ground state of the confinement potential in the $z$ direction.
} 
where $a_\alpha$ and $a_{\ua\da}$ are the $s$-wave scattering lengths between like and unlike bosons, respectively, in the 3D space. 
For simplicity, we set $g_{\ua\ua}=g_{\da\da}\equiv g>0$ and $N_\ua=N_\da$ in the following. 
We further assume that the synthetic magnetic fields $B_\alpha$ are sufficiently high or the interactions are sufficiently weak 
so that the energy scales of the interaction per atom, $|g_{\alpha\beta}| n$, are much smaller than 
the Landau-level spacing $\hbar\omega_\mathrm{c}:= \hbar qB/M$, 
where $n:= N_\ua/A=N_\da/A$ is the density of atoms in each component. 
In this situation, it is legitimate to employ the LLL approximation in which the Hilbert space is restricted to the lowest Landau level \cite{Cooper08,Ho01,Baym05}. 


When the filling factor $\nu \equiv N/N_v$ is sufficiently high $(\nu \gg 1)$, 
the system is well described by the GP mean-field theory. 
In this theory, the GP energy functional $E[\psi_\ua,\psi_\da]$ is introduced by replacing the field operator $\psih_\alpha(\rv)$ 
by the condensate wave function $\psi_\alpha(\rv)$ in the Hamiltonian \eqref{eq:2compHami}; 
then, the functional is minimized under the conditions $\int d^2\rv |\psi_\alpha|^2=N_\alpha$ ($\alpha=\ua,\da$) to determine the ground-state wave functions $\{ \psi_\alpha(\rv)\}$. 
Using the LLL wave functions which have periodic patterns of zeros and are equivalent to the LLL magnetic Bloch states described in Sec.\ \ref{sec2 LLL}, 
Mueller and Ho \cite{Mueller02} have obtained a rich ground-state phase diagram for the parallel-field case, 
which consists of five different vortex-lattice structures as shown in the upper panels of Fig.~\ref{fig:vorlat}. 
Notably, the GP energy functionals for the parallel- and antiparallel-field cases are related to each other as 
$E_{\mathrm{antiparallel}}[\psi_\ua, \psi_\da] = E_{\mathrm{parallel}}[\psi_\ua, \psi_\da^\ast]$ \cite{Furukawa14}. 
This implies that {\it within} the GP theory, the ground-state wave function of one case can be obtained from that of the other 
through the complex conjugation of the spin-$\da$ component.\footnote{
A similar situation arises for the ferromagnetic and antiferromagnetic Heisenberg models on a bipartite lattice, 
whose classical Hamiltonians are related to each other through the spin inversion $\Sv_j\to-\Sv_j$ on one of the two sublattices. 
} 
Therefore, BECs in antiparallel fields also exhibit a rich variety of vortex-lattice structures 
as shown in Fig.\ \ref{fig:vorlat} in the same way as BECs in parallel fields. 

\subsection{Lowest-Landau-level magnetic Bloch states}\label{sec2 LLL}

To describe the excitation properties of a vortex lattice, it is important to choose the basis consistent with the periodicity of the lattice. 
Following Refs.\ \cite{Kwasigroch12,Rashba97,Burkov10}, we utilize the LLL magnetic Bloch states for this purpose. 
Let $\mathbf{a}_1$ and $\mathbf{a}_2$ be the primitive vectors of a vortex lattice as shown in Fig.\ \ref{fig:vorlat}(f). 
These vectors satisfy 
\begin{equation}\label{eq:quan_flux}
  (\av_1\times\av_2)_z=2\pi\ell^2=A/\Nvor,
\end{equation} 
which implies the presence of one vortex in each component per unit cell.  
The reciprocal primitive vectors are then given by
\begin{equation}\label{eq:bv12}
\bv_1=-\ev_z\times\av_2/\ell^2,~~
\bv_2=\ev_z\times\av_1/\ell^2,
\end{equation}
which satisfy $\av_i\cdot\bv_j=2\pi \delta_{ij}~(i,j=1,2)$. 
Using the pseudomomentum for a spin-$\alpha$ particle
\begin{equation}
  \Kv_\alpha = \pv-q\Av_\alpha+q\Bv_\alpha\times\rv=\pv+\epsilon_\alpha \frac{qB}{2}\ev_z\times\rv, 
\end{equation}
we introduce the magnetic translation operator as
$T_\alpha (\sv) = e^{-i \Kv_\alpha \cdot\sv/\hbar}$ \cite{Zak64}. 
We note that the pseudomomentum $\Kv_\alpha=(K_{\alpha,x},K_{\alpha,y})$ satisfies the commutation relation 
$[K_{\alpha,x},K_{\alpha,y}]=-i\epsilon_\alpha \hbar^2/\ell^2$. 
Starting from the most localized symmetric LLL wave function 
$c_0(\mathbf{r})=e^{-\mathbf{r}^2/4\ell^2} / \sqrt{2\pi \ell^2}$, 
we construct a set of LLL wave functions by multiplying two translation operators as
\begin{equation*}
  c_{\mathbf{m} \alpha}(\mathbf{r})
  =T_\alpha(m_1\mathbf{a}_1)  T_\alpha(m_2\mathbf{a}_2) c_0(\mathbf{r}) 
  =\frac{ (-1)^{m_1m_2} }{\sqrt{2\pi\ell^2}} \exp \left[-\frac{1}{4\ell^2}(\mathbf{r}-\mathbf{r_m})^2 - \frac{i\epsilon_\alpha}{2\ell^2} (\mathbf{r} \times \mathbf{r_m})_z\right],
\end{equation*}
where $\mathbf{r_m} = m_1\mathbf{a}_1 + m_2\mathbf{a}_2$ with $\mv=(m_1,m_2)\in\Zbb^2$. 
Here, $T_\alpha(m_1\mathbf{a}_1)$ and $T_\alpha(m_2\mathbf{a}_2)$ commute with each other 
since every unit cell is pierced by one magnetic flux quantum as seen in Eq.~\eqref{eq:quan_flux}; 
this property justifies the application of Bloch's theorem. 
By superposing $c_{\mv\alpha}(\rv)$ for $\Nvor$ possible translations $\mv$ on a torus, we can construct the LLL magnetic Bloch state as \cite{Rashba97}
\begin{equation}\label{eq:mag_Bloch}
  \Psi_{\mathbf{k}\alpha}(\mathbf{r}) = \frac{1}{\sqrt{\Nvor \zeta(\mathbf{k})}} \sum_{\mathbf{m}}c_{\mathbf{m} \alpha}(\mathbf{r})e^{i \mathbf{k}\cdot\mathbf{r_m}} 
\end{equation}
with the normalization factor 
\begin{equation}\label{eq:zeta_k}
 \zeta (\kv) = \sum_{\mv}  (-1)^{m_1 m_2} e^{-\rv_\mv^2/4\ell^2-i\kv\cdot\rv_{\mv}}.
\end{equation}
This state is an eigenstate of $T_\alpha(\av_j)$ with an eigenvalue $e^{-i\kv\cdot\av_j}$. 

The LLL magnetic Bloch state $\Psi_{\kv\alpha}(\rv)$ represents a vortex lattice with a periodic pattern of zeros for any value of 
the wave vector $\kv$.\footnote{
Mueller and Ho \cite{Mueller02} instead use Jacobi's theta function to express a vortex-lattice wave function. 
Such an expression is obtained by performing the Poisson resummation in Eq.\ \eqref{eq:mag_Bloch} for $\Nvor\to\infty$; see \ref{app:mag_Bloch_theta}.
} 
Indeed, by rewriting Eq.\ \eqref{eq:mag_Bloch} as
\begin{equation*}
\begin{split}
 \sqrt{\Nvor \zeta(\kv)} \Psi_{\kv\alpha} (\rv)
 &=\sum_\mv c_{\mv\alpha}^* (\rv) \exp\left[ i \left( -\frac{\epsilon_\alpha}{\ell^2} \ev_z\times\rv +\kv \right) \cdot\rv_\mv \right]\\
 &=\sum_\mv c_{\mv\alpha}^* (\rv) \exp\bigg\{ -\frac{i\epsilon_\alpha}{\ell^2} \left[ \ev_z\times \left( \rv +\epsilon_\alpha \ell^2 \ev_z\times\kv \right)\right] \cdot\rv_\mv \bigg\}
\end{split}
\end{equation*}
and comparing it with the complex conjugate of the Perelomov overcompleteness equation
$\sum_{\mv} (-1)^{m_1+m_2} c_{\mv\alpha} (\rv) =0$ \cite{Perelomov1971}, 
we find that $\Psi_{\kv\alpha}(\rv)$ has zeros at \cite{Burkov10}
\begin{equation}\label{eq:zeros_Bloch}
 \rv=\rv_\nv+\frac12 (\av_1+\av_2) - \epsilon_\alpha \ell^2 \ev_z\times\kv,~~\nv=(n_1,n_2)\in\Zbb^2.
\end{equation} 

When one describes a triangular vortex lattice of a scalar BEC using a LLL magnetic Bloch state, 
the choice of the wave vector $\kv$ is arbitrary 
once the primitive vectors $\av_1$ and $\av_2$ are set appropriately. 
This is because a change in $\kv$ only leads to a translation of zeros as seen in Eq.~\eqref{eq:zeros_Bloch}. 
The vortex lattices of two-component BECs in Fig.~\ref{fig:vorlat} 
can also be described by the LLL magnetic Bloch states $\Psi_{\qv_\alpha,\alpha}(\rv)~(\alpha=\ua,\da)$; 
however, the wave vectors $\qv_\ua$ and $\qv_\da$ have to be chosen in a way consistent 
with the displacement $u_1\av_1+u_2\av_2$ between the components [see Fig.\ \ref{fig:vorlat}(f)]. 
One useful choice is
\begin{equation}\label{eq:qv_uada}
\begin{split}
  &\qv_\ua = +\frac{\epsilon_\ua}{2\ell^2} \ev_z\times (u_1\av_1+u_2\av_2)=\frac{\epsilon_\ua}{2} (-u_2\bv_1+u_1\bv_2),\\
  &\qv_\da = -\frac{\epsilon_\da}{2\ell^2} \ev_z\times (u_1\av_1+u_2\av_2)=\frac{\epsilon_\da}{2} (+u_2\bv_1-u_1\bv_2).
\end{split}
\end{equation}
Here, we displace the spin-$\ua$ component by $\frac12 (u_1\av_1+u_2\av_2)$ and the spin-$\da$ component by $-\frac12 (u_1\av_1+u_2\av_2)$ 
instead of displacing only one of the components. 
This is useful for avoiding zeros of the normalization factor $\zeta(\kv)$ at some high-symmetry points in the first Brillouin zone \cite{Rashba97}.\footnote{
If we set $\qv_\ua=(\bv_1-\bv_2)/2$ and $\qv_\da=\mathbf{0}$ for square lattices, for example, 
we have $\zeta(\qv_\ua)=0$ and Eq.\ \eqref{eq:mag_Bloch} is not well-defined 
unless we factor out a nonanalytic dependence around the point of our concern \cite{Rashba97}. 
}

\subsection{Representation of the Hamiltonian}\label{sec2 Hamiltonian}

Using the magnetic Bloch states \eqref{eq:mag_Bloch}, we expand the field operator as
$\hat{\psi}_{\alpha}(\mathbf{r}) = \sum_{\mathbf{k}} \Psi_{\mathbf{k} \alpha}(\mathbf{r}) b_{\mathbf{k} \alpha}$,  
where $\kv$ runs over the first Brillouin zone, and $b_{\kv\alpha}$ is a bosonic annihilation operator satisfying 
$[b_{\kv\alpha},b_{\kv'\alpha'}^\dagger]=\delta_{\kv\kv'}\delta_{\alpha\alpha'}$. 
Substituting this expansion into the Hamiltonian, we obtain 
\begin{equation}
\begin{split}
  H &= H_\mathrm{kin} + H_{\mathrm{int} } \\
     &=\frac{\hbar\omega_\mathrm{c}}{2} (\hat{N}_{\ua}+\hat{N}_{\da})
  +\frac{1}{2} \sum_{\alpha,\beta} \sum_{\mathbf{k}_1,\mathbf{k}_2,\mathbf{k}_3,\mathbf{k}_4} 
   V_{\alpha\beta}(\mathbf{k}_1,\mathbf{k}_2,\mathbf{k}_3,\mathbf{k}_4) 
   b^{\dag}_{\mathbf{k}_1 \alpha}b^{\dag}_{\mathbf{k}_2 \beta}b_{\mathbf{k}_3 \beta}b_{\mathbf{k}_4 \alpha} ,
\label{2ndQ hamiltonian}
\end{split}
\end{equation}
where $\hbar\omega_\mathrm{c}/2$ is the LLL single-particle zero-point energy and 
$\hat{N}_\alpha=\sum_\kv b^\dag_{\kv\alpha}b_{\kv\alpha}$ is the number operator for the spin-$\alpha$ component. 
The interaction matrix element $V_{\alpha\beta}(\kv_1,\kv_2,\kv_3,\kv_4)$ is given by
\begin{equation}\label{V alpha beta}
 \begin{split}
  V_{\alpha\beta}(\mathbf{k}_1,\mathbf{k}_2,\mathbf{k}_3,\mathbf{k}_4)&=
  g_{\alpha\beta} \int d^2\mathbf{r}~ 
  \Psi^{\ast}_{\mathbf{k}_1\alpha}(\mathbf{r}) \Psi^{\ast}_{\mathbf{k}_2\beta}(\mathbf{r}) 
  \Psi_{\mathbf{k}_3\beta}(\mathbf{r}) \Psi_{\mathbf{k}_4\alpha}(\mathbf{r}).
 \end{split}
\end{equation}
As described in \ref{app: calc}, this matrix element is calculated to be
\begin{equation}\label{eq:V_kkkk}
  V_{\alpha\beta}(\kv_1,\kv_2,\kv_3,\kv_4)
  =\delta_{\kv_1+\kv_2,\kv_3+\kv_4}^\mathrm{P} \frac{g_{\alpha\beta}}{2A} \frac{S_{\alpha\beta} (\kv_1,\kv_2,\kv_3)}{\sqrt{\zeta({\kv_1})\zeta({\kv_2})\zeta({\kv_3})\zeta({\kv_4})}}.
\end{equation}
Here, $\delta_{\kv\kv'}^\mathrm{P}:= \sum_{\Gv} \delta_{\kv,\kv'+\Gv}$ is the periodic Kronecker's delta, where $\Gv$ runs over the reciprocal lattice vectors. 
In the case of parallel fields, the function $S_{\alpha\beta}(\kv_1,\kv_2,\kv_3)$ does not depend on $\alpha$ or $\beta$, and is given by
\begin{equation}\label{eq:S_kkk}
\begin{split}
  S(\kv_1,\kv_2,\kv_3)=&\sum_{\pv\in\{0,1\}^2} (-1)^{p_1p_2}
  e^{-\rv_\pv^2/4\ell^2 + i\kv_3\cdot\rv_\pv} \tilde{\zeta} \left(\kv_1+\kv_2-2\kv_3+(\rv_\pv\times\ev_z-i\rv_\pv)/2\ell^2 \right)\\
  &\times\zeta \left(\kv_1+(\rv_\pv\times\ev_z+i\rv_\pv)/4\ell^2\right) \zeta \left(\kv_2+(\rv_\pv\times\ev_z+i\rv_\pv)/4\ell^2\right) , 
\end{split}
\end{equation}
where
\begin{equation}\label{eq:zetat_k}
  \tilde{\zeta} (\kv) := \sum_\mv e^{-\rv_\mv^2/2\ell^2-i\kv\cdot\rv_\mv}. 
\end{equation}
In the case of antiparallel fields, $S_{\alpha\beta}(\kv_1,\kv_2,\kv_3)$ depends on $\alpha$ and $\beta$, and is given in terms of $S(\kv_1,\kv_2,\kv_3)$ defined above by
\begin{equation}\label{eq:S_kkk^AP}
\begin{split}
  S_{\ua\ua}(\kv_1,\kv_2,\kv_3)=S(\kv_1,\kv_2,\kv_3),~~
  S_{\da\da}(\kv_1,\kv_2,\kv_3)=S(-\kv_1,-\kv_2,-\kv_3)^*,\\
  S_{\ua\da}(\kv_1,\kv_2,\kv_3)=S(\kv_1,-\kv_3,-\kv_2),~~
  S_{\da\ua}(\kv_1,\kv_2,\kv_3)=S(-\kv_1,\kv_3,\kv_2)^*.
\end{split}
\end{equation}

\subsection{Bogoliubov approximation}\label{sec2 Bogoliubov}

At high filling factors, the condensate is only weakly depleted 
and we can apply the Bogoliubov approximation \cite{Pethick_Smith_2008, Sinova02, Matveenko11, Kwasigroch12}.\footnote{
In the thermodynamic limit, however, this approximation is not valid 
since the fraction of quantum depletion diverges as $\frac1N \sum_{\kv\ne\zerov,\alpha} \langle \bt_{\kv\alpha}^\dagger \bt_{\kv\alpha} \rangle \sim \ln(\Nvor)/\nu$ \cite{Sinova02,Kwasigroch12}. 
The Bogoliubov theory is still applicable 
since $\Nvor$ is at most of the order of 100 in typical experiments of ultracold atomic gases \cite{Schweikhard04_1comp}. 
}
Provided that the condensation occurs at the wave vector $\qv_\alpha$ in the spin-$\alpha$ component, it is useful to introduce
\begin{equation}\label{eq:bt_Vt}
  \bt_{\kv\alpha}:=b_{\qv_\alpha+\kv,\alpha},~~ 
  \Vt_{\alpha\beta} (\kv_1,\kv_2,\kv_3,\kv_4):=V_{\alpha\beta}(\qv_\alpha+\kv_1,\qv_\beta+\kv_2,\qv_\beta+\kv_3,\qv_\alpha+\kv_4).
\end{equation}
By setting 
\begin{equation}\label{eq:b_cond}
  \bt_{\zerov \alpha} \simeq \bt_{\zerov \alpha}^\dagger \simeq \sqrt{ N_\alpha-\sum_{\kv\ne\zerov} \bt_{\kv\alpha}^\dagger \bt_{\kv\alpha} }
\end{equation}
and retaining terms up to the second order in $\bt_{\kv\alpha}$ and $\bt_{\kv\alpha}^\dagger$~($\kv\ne 0$), we obtain the following Bogoliubov Hamiltonian: 
\begin{equation}\label{eq:H_bb}
\begin{split}
  H_\mathrm{int} 
  = &\frac12 \sum_{\alpha,\beta} N_\alpha N_\beta\Vt_{\alpha\beta}(\zerov,\zerov,\zerov,\zerov) 
  - \frac12 \sum_{\kv\ne\zerov}  \sum_\alpha \left[ h_\alpha(\kv) + \omega_{\alpha\alpha}(\kv) \right] \\
  &+\frac12 \sum_{\kv\ne\zerov} \left( \bt_{\kv\ua}^\dagger, \bt_{\kv\da}^\dagger, \bt_{-\kv,\ua}, \bt_{-\kv,\da} \right)
  \mathcal{M}(\kv)
 \begin{pmatrix}  \bt_{\kv\ua}\\ \bt_{\kv\da}\\ \bt_{-\kv,\ua}^\dagger\\ \bt_{-\kv,\da}^\dagger \end{pmatrix}.
\end{split}
\end{equation}
Here, the matrix $\mathcal{M} (\kv)$ is given by
\begin{equation}\label{eq:Mmat}
  \mathcal{M} (\kv) =
  \begin{pmatrix}
  h_\ua (\kv)+\omega_{\ua\ua}(\kv) & \omega_{\ua\da}(\kv) & \lambda_{\ua\ua}(\kv) & \lambda_{\ua\da}(\kv) \\
  \omega_{\da\ua}(\kv) & h_\da(\kv)+\omega_{\da\da}(\kv) & \lambda_{\da\ua}(\kv) & \lambda_{\da\da}(\kv) \\
  \lambda_{\ua\ua}^*(\kv) & \lambda_{\da\ua}^*(\kv) & h_\ua(-\kv)+\omega_{\ua\ua}(-\kv) & \omega_{\da\ua}(-\kv) \\
  \lambda_{\ua\da}^*(\kv) & \lambda_{\da\da}^*(\kv) & \omega_{\ua\da}(-\kv) & h_\da(-\kv)+\omega_{\da\da}(-\kv) \\ 
  \end{pmatrix},
\end{equation}
where
\begin{equation}
\begin{split}
  &h_\alpha(\kv) := \sum_\beta N_\beta \left[ \Vt_{\alpha\beta}(\kv,\zerov,\zerov,\kv) - \Vt_{\alpha\beta}(\zerov,\zerov,\zerov,\zerov) \right],\\
  &\omega_{\alpha\beta} (\kv) := \sqrt{N_\alpha N_\beta} \Vt_{\alpha\beta} (\kv,\zerov,\kv,\zerov),~~
  \lambda_{\alpha\beta} (\kv) := \sqrt{N_\alpha N_\beta} \Vt_{\alpha\beta} (\kv,-\kv,\zerov,\zerov).
\end{split}
\end{equation}

To diagonalize the Bogoliubov Hamiltonian \eqref{eq:H_bb}, we perform the Bogoliubov transformation
\begin{equation}\label{eq:b_gamma}
  \begin{pmatrix}  \bt_{\kv\ua}\\ \bt_{\kv\da}\\ \bt_{-\kv,\ua}^\dagger\\ \bt_{-\kv,\da}^\dagger \end{pmatrix}
  = W(\kv) 
  \begin{pmatrix}  \gamma_{\kv,1}\\ \gamma_{\kv,2}\\ \gamma_{-\kv,1}^\dagger\\ \gamma_{-\kv,2}^\dagger \end{pmatrix}, ~
W(\kv)=
 \begin{pmatrix}
 \Ucal(\kv) & \Vcal^*(-\kv) \\ \Vcal(\kv) & \Ucal^*(-\kv)
 \end{pmatrix}. 
\end{equation}
Here, $W(\kv)$ is a paraunitary matrix satisfying 
\begin{equation}\label{eq:WtauW}
  W^\dagger(\kv) \tau_3 W(\kv)=W(\kv)\tau_3 W^\dagger(\kv)=\tau_3 := \mathrm{diag}(1,1,-1,-1),
\end{equation}
which ensures the invariance of the bosonic commutation relation. 
If the matrix $W(\kv)$ is chosen to satisfy 
\begin{equation}\label{eq:WMW}
  W^\dagger (\kv) \mathcal{M}(\kv) W(\kv) = \mathrm{diag} (E_1(\kv),E_2(\kv),E_1(-\kv),E_2(-\kv)),
\end{equation}
the Bogoliubov Hamiltonian is diagonalized as
\begin{equation}\label{eq:H_gg}
 \begin{split}
  H_\mathrm{int} 
  = \frac12 \sum_{\alpha,\beta} N_\alpha N_\beta\Vt_{\alpha\beta}(\zerov,\zerov,\zerov,\zerov) 
  - \frac12 \sum_{\kv\ne\zerov}  \sum_\alpha \left[ h_\alpha(\kv) + \omega_{\alpha\alpha}(\kv) \right] 
  + \sum_{\kv\ne\zerov} \sum_{i=1,2} E_i(\kv) \left( \gamma_{\kv i}^\dagger \gamma_{\kv i}+\frac12 \right) .
 \end{split}
\end{equation}
By multiplying Eq.~\eqref{eq:WMW} from the left by $W(\kv)\tau_3$ and using Eq.\ \eqref{eq:WtauW}, one finds 
\begin{equation}
  \tau_3 \mathcal{M}(\kv) W(\kv) = W(\kv) \mathrm{diag} (E_1(\kv),E_2(\kv),-E_1(-\kv),-E_2(-\kv)). 
\end{equation}
Therefore, the excitation energies $E_i(\kv)~(i=1,2)$ can be obtained as the right eigenvalues of $\tau_3\mathcal{M}(\kv)$.

With the Bogoliubov Hamiltonian \eqref{eq:H_gg}, the field operator shows the following time evolution:
\begin{equation}\label{eq:psi_rt}
 \psih_\alpha(\rv,t)\simeq \sqrt{N_\alpha} \Psi_{\qv_\alpha,\alpha}(\rv) +
 \sum_{\kv\ne \zerov} \Psi_{\qv_\alpha+\kv,\alpha}(\rv)
 \sum_{i=1,2}  \left[\Ucal_{\alpha i}(\kv)e^{-iE_i(\kv)t/\hbar}\gamma_{\kv i}
+\Vcal_{\alpha i}^*(-\kv) e^{iE_i(-\kv)t/\hbar}\gamma_{-\kv, i}^\dagger
  \right].
\end{equation}
If we replace $\gamma_{\kv i}$ and $\gamma_{\kv i}^\dagger$ by c-numbers, 
we may view this equation as the classical time evolution of a condensate wave function $\psi_\alpha(\rv,t)$. 
In particular, by setting $\gamma_{\kv i},\gamma_{\kv i}^\dagger \to c\sqrt{N_\alpha}=c\sqrt{nA}\ne 0$ 
(with $c$ being a real constant) for the specific mode $(\kv,i)$, we obtain
\begin{equation}\label{eq:psi_rt_ki}
 \frac{\psi_\alpha(\rv,t)}{\sqrt{n}} 
 =\sqrt{A} \Psi_{\qv_\alpha,\alpha}(\rv) +
 c\sqrt{A}\left[ 
 \Psi_{\qv_\alpha+\kv,\alpha}(\rv)\Ucal_{\alpha i}(\kv)e^{-iE_i(\kv)t/\hbar}
+\Psi_{\qv_\alpha-\kv,\alpha}(\rv)\Vcal_{\alpha i}^*(\kv) e^{iE_i(\kv)t/\hbar}
  \right].
\end{equation}
This can be used to show how the density profiles $|\psi_\alpha(\rv,t)|^2/n~(\alpha=\ua,\da)$ and the vortex positions change in time in the concerned mode $(\kv,i)$. 
In doing so, it is useful to use the representation of $\sqrt{A} \Psi_{\kv,\alpha}(\rv)$ in terms of Jacobi's theta function [Eq.~\eqref{eq:mag_Bloch_theta} in \ref{app:mag_Bloch_theta}] as this function is supported in various computing systems.\footnote{We used Mathematica and took $W(\kv)$ with the phase choices $\Ucal_{\ua i}(\kv)>0~(i=1,2)$ 
in obtaining the density profiles in Figs.\ \ref{fig:mode_lowE} and \ref{fig:mode_highsym}.} 


\subsection{Numerical results}\label{sec2 results}

\newcommand{\Thetas}{\tilde{\Theta}}
\newcommand{\Tcal}{{\cal T}}
\newcommand{\Tt}{\tilde{T}}
\newcommand{\para}{\mathrm{(P)}}
\newcommand{\apara}{\mathrm{(AP)}}

We use the formulation described above to numerically calculate the Bogoliubov excitation spectrum $\{E_i(\kv)\}$ in the following way. 
For a given wave vector $\kv$, we calculate the matrix $\mathcal{M}(\kv)$ in Eq.~\eqref{eq:Mmat} by using Eqs.\ \eqref{eq:V_kkkk}, \eqref{eq:S_kkk}, and \eqref{eq:S_kkk^AP}. 
We note that each of the functions $\zeta(\cdot)$ and $\tilde{\zeta}(\cdot)$ used in Eq.~\eqref{eq:S_kkk} 
involves an infinite sum but only with respect to two integer variables [see Eqs.~\eqref{eq:zeta_k} and \eqref{eq:zetat_k}], 
which can numerically be taken with high accuracy. 
We then calculate the right eigenvalues of $\tau_3\mathcal{M}(\kv)$ to obtain $\{E_i(\kv)\}$. 

\begin{figure}
 \centering
    \includegraphics[width=16cm]{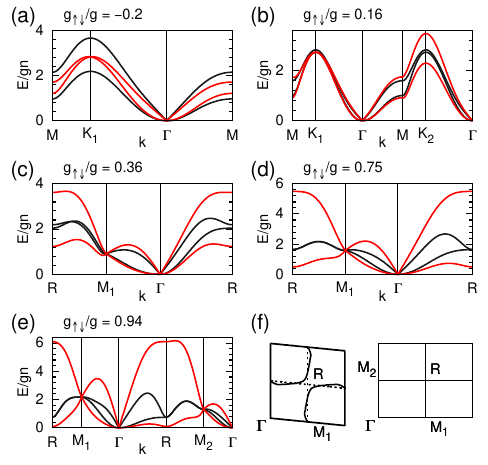} 
 \caption{
Bogoliubov excitation spectra $\{E_i(\kv)\}$ (scaled by $gn$) for the lattice structures shown in Fig.~\ref{fig:vorlat}: 
(a) overlapping triangular, (b) interlaced triangular, (c) rhombic, (d) square, and (e) rectangular lattices. 
Each panel shows both results of parallel (black) and antiparallel (red) magnetic fields. 
Excitation spectra are calculated along the paths indicated by dotted arrows shown in the lower panels of Fig.~\ref{fig:vorlat}. 
The left and right panels in (f) show the lines (solid) in the Brillouin zones 
along which the two bands touch in the cases of (c) rhombic and (e) rectangular lattices, respectively, under parallel fields. 
Dashed straight lines connecting the centers of the edges are guides to the eyes. 
}\label{fig:spectrum}
\end{figure}

\begin{figure}
\centering
\includegraphics[width=14cm]{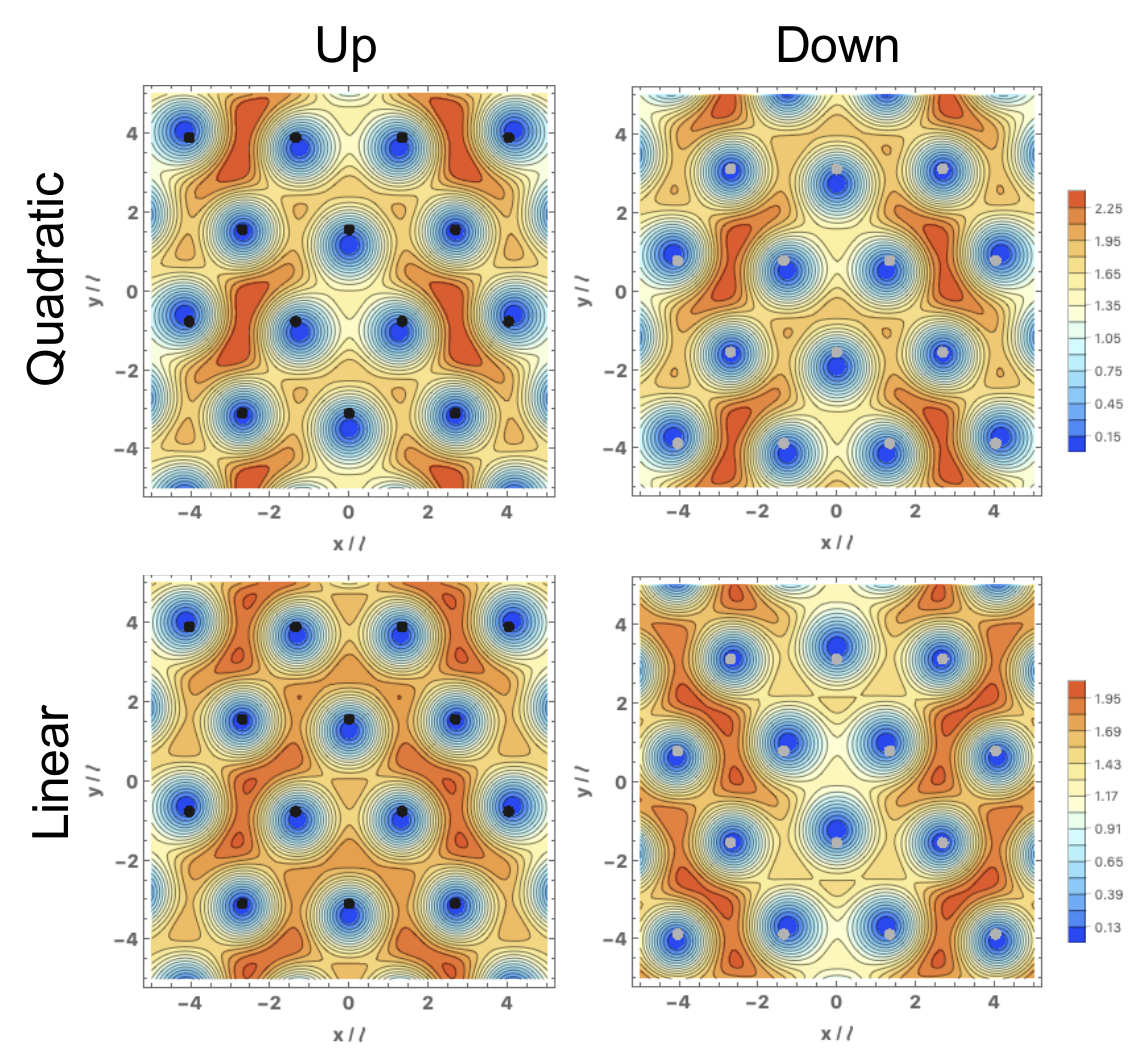}
\caption{
Density profiles $|\psi_\alpha(\rv,t=0)|^2/n~(\alpha=\ua,\da)$
of the modes with quadratic ($i=2$) and linear ($i=1$) dispersion relations at $\kv=(0.2a/\ell^2,0)$ 
for interlaced triangular lattices in parallel fields. 
Calculations were performed using Eq.\ \eqref{eq:psi_rt_ki} with $c=0.3$. 
A relatively large value of $c$, which might be beyond the scope of the Bogoliubov theory, 
is taken to emphasize the changes due to the excitations. 
Black (gray) circles indicate the locations of spin-$\ua$ ($\da$) vortices in the ground state. 
}\label{fig:mode_lowE}
\end{figure}

Figure \ref{fig:spectrum} presents the obtained energy spectra for all the lattice structures in Fig.~\ref{fig:vorlat} 
and for both the parallel- and antiparallel-field cases. 
In all the cases, we find that there appear two modes with linear and quadratic dispersion relations at low energies around the $\Gamma$ point. 
Furthermore, we find anisotropy of the coefficients of these dispersion relations. 
For example, such anisotropy can clearly be seen along the path $M_1 \to \Gamma \to R$ 
for (c) rhombic, (d) square, and (e) rectangular lattices. 
We discuss such anisotropy in detail in later sections. 

To gain some physical insight into the low-energy excitation modes, we present in Fig.\ \ref{fig:mode_lowE} 
the density profiles of the modes with quadratic ($i=2$) and linear ($i=1$) dispersion relations at $\kv=(0.1a/\ell^2,0)$ 
for (b) interlaced triangular lattices in parallel fields. 
As seen in this figure, vortices move perpendicularly to $\kv$ relative to the ground state. 
Furthermore, spin-$\ua$ and $\da$ vortices show in-phase (anti-phase) oscillations in the $i=2$ ($i=1$) mode. 
Specifically, around $k_x=0$, both spin-$\ua$ and $\da$ vortices move in the $-y$ direction in the $i=2$ mode (upper panels of Fig.\ \ref{fig:mode_lowE}) 
while they move in opposite directions ($\mp y$) in the $i=1$ mode (lower panels). 
Similar results are also obtained in the antiparallel-field case (not shown). 
These features are consistent with those obtained from the effective field theory described in Sec.\ \ref{sec:Effect}. 

Apart from the low-energy features, the spectra in Fig.\ \ref{fig:spectrum} also exhibit unique structures of band touching at some high-symmetry points or along lines in the Brillouin zone. 
In particular, the spectra for (c) rhombic, (d) square, and (e) rectangular lattices in parallel fields exhibit line nodes, whose locations in the Brillouin zones are shown in Fig.\ \ref{fig:spectrum}(f). 
This can be understood as a consequence of a ``fractional'' translation symmetry\footnote{
We give more precise definitions of the fractional translation operators $\Tcal^\para$ and $\Tcal^\apara$ in \ref{app:fractrans}. 
} \cite{Young15,Parameswaran13}. 
Namely, in these cases, the system is invariant under the product $\Tcal^\para$ of the translation by $\av_3/2$ 
and the spin reversal $\ua\leftrightarrow\da$, where $\av_3:=\av_1+\av_2$. 
Since the unitary operator $\Tcal^\para$ commutes with the Bogoliubov Hamiltonian and $\left( \Tcal^\para \right)^2$ gives the translation by $\av_3$, 
the Bloch states at $\kv$ can be chosen to be the eigenstates of $\Tcal^\para$ with 
$\Tcal^\para\ket{w_\kv^\pm}=\pm e^{-i\kv\cdot\av_3/2}\ket{w_\kv^\pm}$. 
For a smooth change $\kv\to \kv+\bv_i~(i=1,2)$, the two eigenstates must switch places, indicating the occurrence of an odd number of degeneracies. 
In Fig.\ \ref{fig:spectrum}(f), we can indeed confirm that starting from any point other than the line nodes, the degeneracy occurs once or three times for the above changes of $\kv$. 
The emergence of point nodes at the $M_1$ and $M_2$ points for the same lattices [(c), (d), and (e)] in antiparallel fields 
can be understood by considering the symmetry under the product $\Tcal^\apara$ of the time reversal and the translation by $\av_3/2$. 
Since $\left( \Tcal^\apara \right)^2$ is equal to the translation by $\av_3$, 
we have $(\Tcal^\apara)^2=e^{-i\kv\cdot\av_3}$ in the subspace with the wave vector $\kv$. 
The Kramers degeneracy thus occurs at time-reversal-invariant momenta with $e^{-i\kv\cdot\av_3}\ne 1$, which is the case for $\kv=\bv_1/2$ and $\bv_2/2$ ($M_1$ and $M_2$ points). 
In \ref{app: element}, we further discuss some other features of the spectra at high-symmetry points, 
such as the coincidence of the excitation energies between the two types of fields at the $M_1$ and $M_2$ points in Fig.\ \ref{fig:spectrum}(c), (d), and (e) 
by using the numerical data of the Bogoliubov Hamiltonian matrix $\mathcal{M} (\kv)$ and the density profiles of the excitation modes. 


\section{Effective field theory for low-energy excitation spectra}\label{sec:Effect}

We have seen in the preceding section that vortex lattices of two-component BECs exhibit two excitation modes with linear and quadratic dispersion relations at low energies. 
Here we derive such low-energy dispersion relations by using an effective field theory. 
Specifically, we apply the formalism for a scalar BEC developed by Watanabe and Murayama \cite{WatanabeMurayama13} to the present two-component case. 
This approach is equivalent to the hydrodynamic theory applied by Ke\c{c}eli and Oktel \cite{Oktel06} to two-component BECs in parallel fields. 
However, we point out that an important term is missing in the elastic energy of vortex lattices used in Ref.~\cite{Oktel06}. 
This term is crucial for explaining the anisotropy of the quadratic dispersion relation for interlaced triangular lattices. 
Furthermore, we derive remarkable ``rescaling'' relations between the spectra for the two types of synthetic fields; 
these relations are confirmed for overlapping triangular lattices in Sec.\ \ref{sec:aniso}. 

\subsection{Effective Lagrangian for phase variables}

The Lagrangian density of the two-component BECs corresponding to the Hamiltonian \eqref{eq:2compHami} is given by \cite{Pethick_Smith_2008}
\begin{equation}\label{eq:2comp_Lag}
 \mathcal{L} = \sum_{\alpha} \left[ 
  \frac{i\hbar}{2}(\psi_{\alpha}^{\dag} \dot{\psi}_{\alpha} - \dot{\psi}_\alpha^\dag \psi_\alpha)
  - \frac{1}{2M} |(-i \hbar \nabla- q \mathbf{A}_{\alpha} )\psi_{\alpha}|^2 \right]
  -  \sum_{\alpha,\beta} \frac{g_{\alpha\beta}}{2} |\psi_\alpha|^2|\psi_\beta|^2,
\end{equation}
where $\psi_{\alpha}(\rv,t)$ is the bosonic field for the spin-$\alpha$ component. 
To describe the low-energy properties of the BECs, it is useful to decompose the field as
$\psi_{\alpha} = \sqrt{n_{\alpha}} \exp(- i \theta_{\alpha})$, 
where $n_{\alpha} (\rv,t)$ and $\theta_{\alpha}(\rv,t)$ are the density and phase variables, respectively. 
Substituting this into Eq.~\eqref{eq:2comp_Lag} and keeping only the leading terms in the derivative expansion, we obtain
\begin{equation}\label{eq:2comp_BEC-Lag}
  \mathcal{L} = \mu_{\ua} n_{\ua} + \mu_{\da} n_{\da}- \frac{g}{2}(n_{\ua}^2+n_{\da}^2) - g_{\uparrow\downarrow}n_{\ua}n_{\da} , 
\end{equation}
where
\begin{equation}\label{eq:effective chemical potential}
\mu_{\alpha} = \hbar\dot{\theta}_{\alpha} - \frac{1}{2M}(\hbar\nabla\theta_{\alpha} + q \mathbf{A}_{\alpha} )^2  
\end{equation}
is an effective chemical potential for the spin-$\alpha$ component. 
Introducing $n_\pm := n_\ua \pm n_\da$ and $g_{\pm}:= g \pm g_{\uparrow\downarrow}$, we can rewrite Eq.~\eqref{eq:2comp_BEC-Lag} as
\begin{equation}\label{eq:2comp_BEC-Lag-spin_charge}
\mathcal{L}  = -\frac{g_+}4 n_+^2 -\frac{g_-}4 n_-^2 + \frac{\mu_\ua+\mu_\da}2 n_+ + \frac{\mu_\ua-\mu_\da}{2} n_- . 
\end{equation}
By integrating out $n_\pm(\rv,t)$, we obtain the effective Lagrangian for the phase variables $\{\theta_\alpha(\rv,t)\}$ as
\begin{equation}\label{eq:L_mu_pm}
\mathcal{L} =\frac{1}{4g_+}(\mu_\ua+\mu_\da)^2+ \frac{1}{4g_-}(\mu_\ua-\mu_\da)^2 .
\end{equation}

\subsection{Relation between vortex displacement and phase variables} 

In the presence of vortices, the phase variables $\{ \theta_\alpha(\rv,t) \}$ involve singularities. 
It is thus useful to decompose $\theta_{\alpha}$ into regular and singular parts as 
$\theta_{\alpha} =\theta_{\mathrm{reg},\alpha} + \theta_{\mathrm{sing},\alpha}$. 
Since the singular part $\theta_{\mathrm{sing},\alpha}$ varies rapidly in space, 
it is not a convenient variable for a coarse-grained description over long length scales. 
To describe the long-wavelength physics, it is useful to start from the vortex-lattice ground state (as in Fig.\ \ref{fig:vorlat}) 
and to consider small displacement of vortices from the equilibrium positions. 
Specifically, we introduce the displacement vector field 
$\mathbf{u}_{\alpha}(\mathbf{r},t)=\mathbf{r} - \mathbf{X}_{\alpha}(\mathbf{r},t)$,
where $\mathbf{r} $ is the equilibrium position of the vortex and $\mathbf{X}_{\alpha}$ is the position at time $t$. 
The derivatives of the singular part $\theta_{\mathrm{sing},\alpha}$ of the phase are related 
to the displacement $\uv_\alpha$ as \cite{WatanabeMurayama13}
\begin{equation*}
 \hbar\dot{\theta}_{\sing,\alpha}  =  -\frac{qB_\alpha}{2} (\uv_\alpha \times\dot{\uv}_\alpha)_z,~~
 \hbar\nabla \theta_{\sing,\alpha} +q\Av_\alpha =qB_\alpha \ev_z\times\uv_\alpha -\frac{qB_\alpha}{2} \sum_{i,j}\epsilon_{ij}u^i_\alpha \nabla u^j_\alpha,
\end{equation*}
where $\epsilon_{ij}$ is an antisymmetric tensor with $\epsilon_{xy}=-\epsilon_{yx}=+1$. 
The effective chemical potential in Eq.\ (\ref{eq:effective chemical potential}) can then be expressed 
in terms of $\{\theta_{\mathrm{reg},\alpha}, \uv_\alpha\}$ as
\begin{equation*}
 \mu_{\alpha} = \hbar\dot{\theta}_{\reg,\alpha} - \frac{qB_{\alpha}}{2} ( \mathbf{u}_{\alpha} \times \dot{\mathbf{u}}_{\alpha} )_z
 - \frac{1}{2M} \left( \hbar\nabla \theta_{\mathrm{reg},\alpha} + qB_{\alpha} \mathbf{e}_z \times \mathbf{u}_\alpha - \frac{qB_{\alpha}}{2} \sum_{ij}\epsilon_{ij} u_{\alpha}^i\nabla u_{\alpha}^j \right)^2.
\end{equation*}
One should also note that the displacement $\uv_\alpha(\rv,t)$ leads to a change in the elastic energy $\int d^2\rv ~\Ecal_\elastic (\uv_\alpha,\partial_i\uv_\alpha)$. 
Here, the form of the elastic energy density $\Ecal_\elastic$ depends on the type of a lattice as discussed in the next section and \ref{app:elastic}. 
The effective Lagrangian in terms of $\{\theta_{\mathrm{reg},\alpha},\uv_\alpha\}$ is then obtained as
\begin{equation}\label{eq:Leff_mu_pm}
 \mathcal{L}_\eff= \frac{1}{4g_+}(\mu_\ua+\mu_\da)^2+ \frac{1}{4g_-}(\mu_\ua-\mu_\da)^2 
 - \Ecal_\elastic.
\end{equation}
Here, the difference from Eq.\ \eqref{eq:L_mu_pm} occurs because the rapidly varying $\{\theta_{\mathrm{sing},\alpha}\}$ 
have been replaced by the slowly varying $\{\uv_\alpha\}$ via coarse graining. 

The ground state of $H-\mu_0 (N_\ua+N_\da)$ is given by $\theta_\reg=\mu_0 t/\hbar$ and $\uv_\alpha=\zerov$. 
To discuss the low-energy properties, it is therefore useful to introduce 
$\varphi_{\alpha} = \mu_0 t/\hbar -\theta_{\mathrm{reg},\alpha}$ 
and expand the Lagrangian (\ref{eq:Leff_mu_pm}) in terms of $\{ \varphi_\alpha, \uv_\alpha\}$. 
Keeping only the quadratic terms in these variables, we obtain
\begin{equation} \label{eq:L_phi_u}
 \mathcal{L}_\eff = \frac{\hbar^2\dot{\varphi}_+^2}{4 g_+ } + \frac{\hbar^2\dot{\varphi}^2_-}{4 g_- } 
  -\frac{\mu_0}{g_+} \sum_\alpha \left[ \frac{qB_{\alpha}}{2} (\mathbf{u}_{\alpha} \times \dot{\mathbf{u}}_{\alpha})_z
 +\frac{1}{2M}(\hbar\ev_z\times\nabla\varphi_\alpha+qB_\alpha \uv_\alpha)^2 \right] 
 - \Ecal_\elastic , 
\end{equation}
where $\varphi_{\pm} := \varphi_1 \pm \varphi_2$. 
Because $\{\uv_\alpha\}$ have the mass term $-\uv_\alpha^2$, one can expect that they can safely be integrated out in the discussion of low-energy dynamics. 
To do so, it is useful to derive the Euler-Lagrange equations for $\{\uv_\alpha\}$: 
\begin{equation}\label{eq:EOM_u}
 \uv_\alpha+\epsilon_\alpha \ell^2 \ev_z\times\nabla\varphi_\alpha
 -\frac{\epsilon_\alpha}{\omega_\mathrm{c}} \ev_z\times\dot{\uv}
 +\frac{g_+\ell^2}{\mu_0\hbar\omega_\mathrm{c}} \left[ \frac{\partial\Ecal_\elastic}{\partial \uv_\alpha} 
 -\sum_j \partial_j \left( \frac{\partial\Ecal_\elastic}{\partial\left(\partial_j\uv_\alpha\right)} \right)\right] =0,
\end{equation}
where we use the cyclotron frequency $\omega_\mathrm{c}=qB/M$ and the magnetic length $\ell=\sqrt{\hbar/qB}$. 
The third and fourth terms on the left-hand side can be ignored in the LLL approximation 
($\hbar\omega,|g_{\alpha\beta}| n \ll \hbar\omega_\mathrm{c}$, where $\omega$ is the frequency of our interest). 
Similar relations are also found in hydrodynamic theory \cite{Sonin76, Williams77, Baym83_1,Chandler86_2,Sonin87,Baym03, Cozzini04, Sonin05_cont,Oktel06}. 
Introducing $ \mathbf{u}_{\pm} := \mathbf{u}_{\ua} \pm \mathbf{u}_{\da} $, Eq.\ \eqref{eq:EOM_u} can be rewritten as
\begin{equation}\label{eq:EOM para/anti}
\begin{split}
\mathbf{u}_{\pm} = 
 \left\{
 \begin{aligned}
 &-\ell^2 \mathbf{e}_z \times \nabla \varphi_{\pm}  \qquad \text{(parallel fields)} ; \\
 &-\ell^2 \mathbf{e}_z \times \nabla \varphi_{\mp}  \qquad \text{(antiparallel fields)} .
 \end{aligned}
 \right.
 \qquad \text{}
\end{split}
\end{equation}
These relations indicate that the vortex displacements $\uv_\pm$ and the phases $\varphi_\pm$ 
are coupled in an opposite manner between the parallel- and antiparallel-field cases. 
Namely, the symmetric $\mathbf{u}_{+}$ (antisymmetric $\mathbf{u}_{-}$) is coupled to the symmetric $\varphi_{+}$ (antisymmetric $\varphi_{-}$) in parallel fields, 
while they are coupled in a crossed manner in antiparallel fields.
Equation \eqref{eq:EOM para/anti} also indicates that the vortex displacement is perpendicular to the wave vector $\kv$, 
which is consistent with the results shown in Fig.\ \ref{fig:mode_lowE}. 

Substituting Eq.~\eqref{eq:EOM para/anti} into the Lagrangian density \eqref{eq:L_phi_u}, 
we obtain the Lagrangian density in terms of $\{\varphi_\pm\}$, 
which can be used to determine the excitation spectrum. 
For this purpose, we need to determine the form of the elastic energy density $\Ecal_\elastic$, which is done next. 

\subsection{Elastic energy}\label{sec:Effect_elastic}

Since the elastic energy is invariant under a uniform change in $\uv_+(\rv,t)$ (i.e., translation of the lattices), 
$\Ecal_\elastic$ should be a function of $\partial_i \uv_+~(i=x,y)$ and $\uv_-$ to the leading order in the derivative expansion. 
We therefore introduce the form
\begin{equation}\label{eq: eff energy elastic}
 \Ecal_\elastic = \Ecal_\elastic^{(+)} (\partial_i \uv_+) + \Ecal_\elastic^{(-)} (\uv_-) + \Ecal_\elastic^{(+-)} (\partial_i\uv_+,\uv_-). 
\end{equation}
To express $\Ecal_\elastic^{(+)}$, it is useful to introduce
\begin{equation}
 w_0:=\partial_x u_+^x+\partial_y u_+^y,~~
 w_1:=\partial_x u_+^x -\partial_y u_+^y,~~
 w_2:=\partial_y u_+^x+\partial_x u_+^y. 
\end{equation}
In the LLL regime, the vortex density stays constant, and therefore $w_0=0$; 
this can also be confirmed by using Eq.~\eqref{eq:EOM para/anti}. 
From a symmetry consideration (see \ref{app:elastic}), each term in Eq.~\eqref{eq: eff energy elastic} can be expressed as
\begin{equation}\label{eq : elastic energy}
\begin{split}
 &\Ecal_\elastic^{(+)} (\partial_i \uv_+) 
 = \frac{gn^2}{2} \left( C_1 w_1{}^2 + C_2 w_2{}^2 + C_3 w_1w_2 \right) , \\
 &\Ecal_\elastic^{(-)} (\uv_-) 
 = \frac{gn^2}{2\ell^2} [D_1 (u_-^x)^2+ D_2 (u_-^y)^2+D_3 u_-^xu_-^y], \\
 &\Ecal_\elastic^{(+-)} (\partial_i \uv_+,\uv_-) 
 = \frac{gn^2}{2\ell} F_1 (w_1u_-^y + w_2u_-^x ),
\end{split}
\end{equation}
where $n:=N_\ua/A=N_\da/A$ is the average number density of each component.
For each of the vortex lattices in Fig.\ \ref{fig:vorlat}(a)-(e), 
the dimensionless elastic constants $\{C_1, C_2, C_3, D_1, D_2,D_3,F_1\}$ satisfy
\begin{equation}\label{eq : parameter}
\begin{split}
  &\text{(a)}~ C_1=C_2\equiv C>0,~ D_1=D_2\equiv D>0,~ C_3=D_3=F_1=0;\\
  &\text{(b)}~ C_1=C_2\equiv C>0, D_1=D_2\equiv D>0,~C_3=D_3=0,~F_1 \ne 0;\\
  &\text{(c)}~ C_1, C_2, D_1,D_2>0,~ C_3,D_3\ne 0,~ F_1=0;\\
  &\text{(d)}~ C_1, C_2>0,~ D_1=D_2\equiv D>0,~C_3=D_3=F_1=0;\\
  &\text{(e)}~ C_1, C_2>0,~ D_1,D_2>0,~C_3=D_3=F_1=0.
\end{split}
\end{equation}
Ke\c{c}eli and Oktel \cite{Oktel06} have considered an elastic energy consisting of $\Ecal_\elastic^{(\pm)}$ above, but have not included $\Ecal_\elastic^{(+-)}$. 
Therefore, in their work, the symmetric and antisymmetric displacements $\uv_\pm$ were decoupled from each other in collective modes. 
In our analysis in \ref{app:elastic}, $\Ecal_\elastic^{(+-)}$ is found to be allowed by symmetry for interlaced triangular lattices. 
As shown below, this part crucially changes the low-energy spectrum, 
and explains the anisotropy of the spectrum for the concerned lattices. 

We note that within the mean-field theory, the elastic energy density $\Ecal_\elastic$ should take the same form [Eqs.~\eqref{eq: eff energy elastic} and \eqref{eq : elastic energy}] 
for the parallel- and antiparallel-field cases because of the exact correspondence of the GP energy functionals between the two cases \cite{Furukawa14}. 
The dimensionless elastic constants are also expected to take the same values between the two cases. 
However, as we will see in Sec.~\ref{sec:aniso}, the elastic constants 
estimated from the numerical results of the energy spectra are different between the two cases. 
We discuss this puzzling issue in Sec.~\ref{sec:aniso_others}. 

\subsection{Excitation spectrum}

The Lagrangian density in terms of $\varphi_\pm$ is obtained 
by substituting Eq.\ \eqref{eq:EOM para/anti} into Eq.\ \eqref{eq:L_phi_u} and using the above $\Ecal_\elastic$. 
After performing the Fourier transformation 
$\varphi_\pm(\rv,t)=\sum_\kv \int \frac{d\omega}{2\pi\sqrt{A}} e^{i(\kv\cdot\rv-\omega t)} \varphi_\pm(\kv,\omega)$, 
we obtain the action
\begin{equation}\label{eq:action_varphi}
\begin{split}
 S = \sum_\kv \int \frac{d\omega}{2\pi} \frac12 \left( \varphi_+(-\kv,-\omega),\varphi_-(-\kv,-\omega) \right) 
 i G (\kv,\omega)^{-1}
 \begin{pmatrix} \varphi_+(\kv,\omega) \\ \varphi_-(\kv,\omega) \end{pmatrix} ,
\end{split}
\end{equation}
where 
\begin{equation}\label{eq:Jmat}
  iG (\kv,\omega)^{-1} =
  \begin{pmatrix}
   \frac{\hbar^2\omega^2}{2g_+} - \Gamma_\pm (\kv) & \pm i \Gamma (\kv) \\
   \mp i  \Gamma(\kv) & \frac{\hbar^2\omega^2}{2g_-} -  \Gamma_\mp (\kv) \\
  \end{pmatrix}
\end{equation}
is the inverse of Green's function in Fourier space with 
\begin{equation}\label{eq:Gamma}
\begin{split}
 \Gamma_+ (\kv) &= gn^2\ell^4 [ C_1(2k_xk_y)^2+ C_2(k_x^2-k_y^2)^2 - C_3 (2k_xk_y)(k_x^2-k_y^2) ],\\
 \Gamma_- (\kv) &= gn^2\ell^2 ( D_1k_y^2+ D_2k_x^2- D_3 k_xk_y ),\\
 \Gamma (\kv) &= \frac12 gn^2\ell^3 F_1 (3k_x^2k_y-k_y^3).
\end{split}
\end{equation}
In Eq.~\eqref{eq:Jmat} [and Eqs.\ \eqref{eq:Ei_Gamma}, \eqref{eq: dispersion}, and \eqref{eq:f-effective} below], 
the upper and lower of the double signs correspond to the parallel- and antiparallel-field cases, respectively. 

The excitation spectrum corresponds to the poles of the Green's function, 
and can thus be obtained by solving the equation $\det [iG(\kv,\omega)^{-1}]=0$. 
Since $\Gamma_-(\kv)\gg \Gamma(\kv)\gg \Gamma_+(\kv)$ for $k\ell\ll 1$, we obtain the low-energy dispersion relations as
\begin{equation}\label{eq:Ei_Gamma}
 E_2(\kv) = \sqrt{2g_\pm \left( \Gamma_+(\kv) - \frac{\Gamma(\kv)^2}{\Gamma_-(\kv)} \right)},~~
 E_1(\kv) = \sqrt{2g_\mp \Gamma_-(\kv)}.
\end{equation}
Using Eq.~\eqref{eq:Gamma} and the fact that $\Gamma_-(\kv)$ is isotropic when $F_1\ne 0$ [see Eq.~\eqref{eq : parameter}], 
we obtain the following explicit expressions 
\begin{equation}\label{eq: dispersion}
\begin{split}
 \frac{E_2 (\kv)}{\sqrt{2}gn} =& \left(\frac{g_\pm}{g}\right)^{\frac12} \ell^2 
     \left[ C_1(2k_xk_y)^2 + C_2(k_x^2-k_y^2)^2 - C_3(2k_xk_y)(k_x^2-k_y^2) -  C_4 \frac{(3k_x^2k_y-k_y^3)^2}{k^2} \right]^{\frac12},\\
 \frac{E_1 (\kv)}{\sqrt{2}gn} =& \left(\frac{g_\mp}{g}\right)^{\frac12} \ell  (D_1k_y^2+D_2k_x^2 - D_3k_xk_y)^{1/2}
\end{split}
\end{equation}
with $C_4:=F_1{}^2/4D_1$. 
We thus find the emergence of quadratic and linear dispersion relations 
whose anisotropy reflects the symmetry of each lattice structure. 
Furthermore, we find that the modes with the quadratic and linear dispersion relations 
originate mainly from the symmetric and antisymmetric parts $\uv_\pm$ of the vortex displacement, respectively 
(we however note that these two parts are mixed slightly in the case of interlaced triangular lattices owing to $F_1\ne 0$). 
This explains the in-phase (anti-phase) oscillations of the $i=2$ ($i=1$) mode found in Fig.\ \ref{fig:mode_lowE}




To discuss the anisotropy further, we parametrize the wave vector in terms of polar coordinates as $\kv=k(\cos\theta, \sin\theta)~(k\ell\ll 1)$ 
and introduce the dimensionless functions $\{f_i(\theta)\}$ via
\begin{equation}\label{eq:excit-spect}  
  E_i(\kv)=\sqrt{2} gn (k\ell)^i f_i(\theta),~~i=1,2.
\end{equation}
Using the dispersion relations \eqref{eq: dispersion} obtained from the effective field theory, these functions are calculated as
\begin{equation}\label{eq:f-effective}  
 \begin{split}
  &f_2 (\theta) = \sqrt{\frac{g_\pm}{g}} [
  C_1\sin^2 (2\theta) +C_2\cos^2 (2\theta) -C_3\sin(2\theta)\cos(2\theta) -C_4\sin^2 (3\theta)]^{1/2} ,\\
  &f_1 (\theta) = \sqrt{\frac{g_\mp}{g}} (D_1\sin^2\theta+D_2\cos^2\theta - D_3\sin\theta\cos\theta)^{1/2}.
 \end{split}
\end{equation}
In this result [and also in Eqs.\ \eqref{eq:Ei_Gamma} and \eqref{eq: dispersion}], 
the dependence on the type of synthetic fields occurs only in the coefficients $\sqrt{g_\pm/g}$. 
This observation leads to the following remarkable relations: 
\begin{equation}\label{eq:P-AP relation} 
  f_2^\mathrm{P}(\theta)\sqrt{\frac{g}{g_+}}= f_2^\mathrm{AP}(\theta) \sqrt{\frac{g}{g_-}} ,~~ 
  f_1^\mathrm{P}(\theta)\sqrt{\frac{g}{g_-}} = f_1^\mathrm{AP}(\theta)\sqrt{\frac{g}{g_+}}  , 
\end{equation}
where the superscripts P and AP refer to the parallel- and antiparallel-field cases, respectively. 
Namely, the functions $\{f_i^\mathrm{P/AP}(\theta)\}$ for the two types of synthetic fields are related to each other by simple rescaling. 
While these rescaling relations are expected for all the lattice structures within the effective field theory, 
we show in the next section that the relations hold only for overlapping triangular lattices and break down for the other lattices.

\section{Anisotropy of low-energy excitation spectra}\label{sec:aniso}

We have seen in Sec.\ \ref{sec2 results} that the Bogoliubov excitation spectrum 
exhibits linear and quadratic dispersion relations at low energies with significant anisotropy in some cases. 
In this section, we analyze this anisotropy further 
by calculating the dimensionless functions $\{f_i (\theta)\}$ defined in Eq.\ \eqref{eq:excit-spect} 
for the cases shown in Fig.~\ref{fig:spectrum}. 
We compare the numerical results with the analytical expressions \eqref{eq:f-effective} obtained by the effective field theory. 
We also examine whether the numerical results satisfy the rescaling relations \eqref{eq:P-AP relation} derived by the effective field theory. 


\subsection{Overlapping triangular lattices}

For (a) overlapping triangular lattices, by using Eqs.~\eqref{eq : parameter} and \eqref{eq:f-effective},
the analytic expressions of $\{f_i^\mathrm{P/AP}(\theta)\}$ for parallel (P) and antiparallel (AP) fields are obtained as
\begin{equation}\label{eq: triangular f_12}
\begin{split}
  f_2^\mathrm{P}(\theta)= \sqrt{\frac{g_+}{g} C},~~ f_1^\mathrm{P}(\theta)= \sqrt{\frac{g_-}{g} D},~~ 
  f_2^\mathrm{AP}(\theta)= \sqrt{\frac{g_-}{g} C}, ~~ f_1^\mathrm{AP}(\theta)= \sqrt{\frac{g_-}{g} D}.
\end{split}
\end{equation}
Notably, these functions show no dependence on $\theta$ in the effective field theory. 

\begin{figure}
 \centering
   \includegraphics[width=15cm, angle=0]{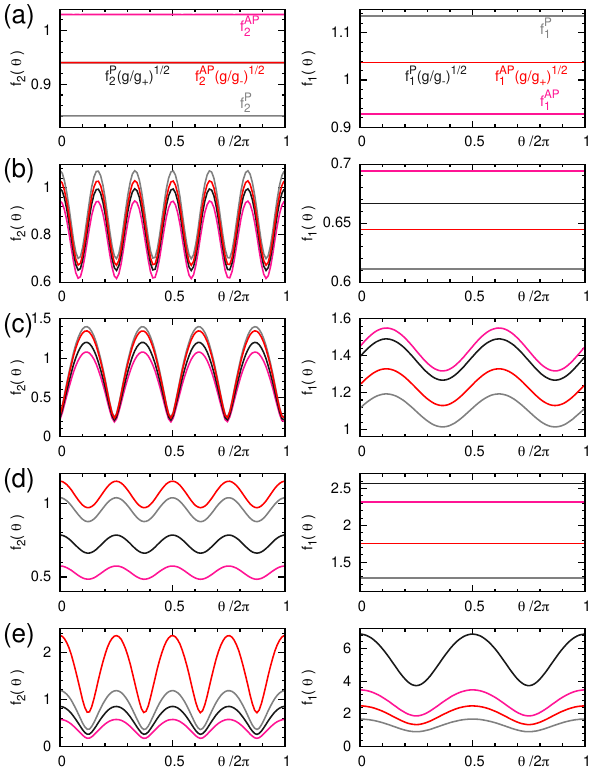} 
  \caption{
Dimensionless functions $f_2^\mathrm{P/AP}(\theta)$ (left) and $f_1^\mathrm{P/AP}(\theta)$ (right) 
for parallel (P; gray) and antiparallel (AP; pink) fields for the same cases as in Fig.\ \ref{fig:spectrum}(a)-(e). 
These functions are calculated from the Bogoliubov excitation spectra $\{E_i(\kv)\}$ 
along a circular path $\kv = k(\cos\theta, \sin\theta)$ with $k=0.001a/\ell^2$ and $\theta\in [0,2\pi)$. 
The rescaled functions [left- and right-hand sides of Eq.\ \eqref{eq:P-AP relation}] are also shown (black and red), 
confirming the rescaling relations \eqref{eq: triangular f_12} through the overlap of the curves for (a) overlapping triangular lattices. 
 }
  \label{fig:postitive_azimuth}
\end{figure}

\begin{figure}
 \centering
   \includegraphics[width=15cm]{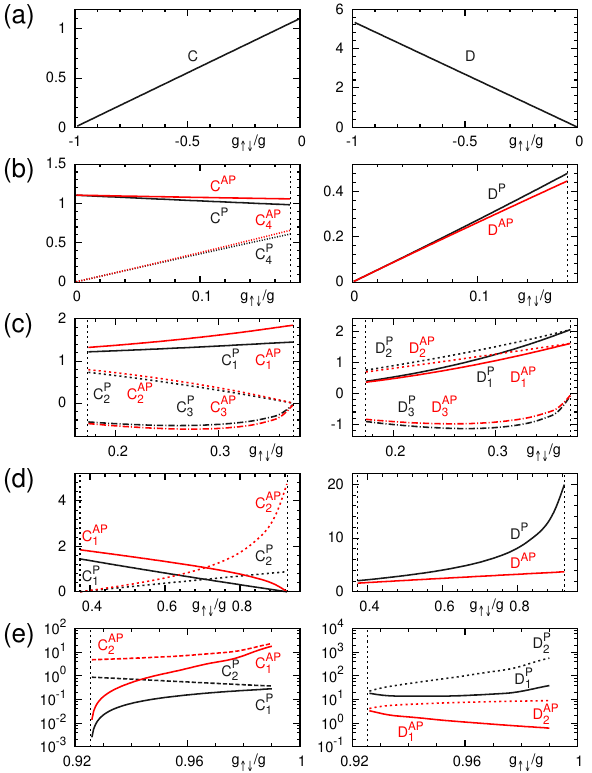} 
  \caption{
Constants $C_i^\mathrm{P/AP}~(i=1,2,3,4)$ (left) and $D_i^\mathrm{P/AP}~(i=1,2,3)$ (right)
for parallel (P; black) and antiparallel (AP; red) fields 
for (a) overlapping triangular, (b) interlaced triangular, (c) rhombic, (d) square, and (e) rectangular lattices 
[see Eq.~\eqref{eq : parameter} for the symmetry constraints on the constants]. 
These are obtained by fitting the numerically obtained functions $\{f_i^\mathrm{P/AP}(\theta)\}$ 
(as in Fig.\ \ref{fig:postitive_azimuth}) using Eq.\ \eqref{eq:f-effective}.  
Semi-logarithmic scales are used in (e). 
Vertical dashed lines indicate the transition points. 
 }
  \label{fig:elastic constants}
\end{figure}

In numerical calculations, we obtain $\{f_i^\mathrm{P/AP}(\theta)\}$ from the data of the Bogoliubov excitation spectra 
along a circular path $\kv=k(\cos\theta, \sin\theta)$ with sufficiently small $k$ and arbitrary $\theta\in[0,2\pi)$. 
Figure~\ref{fig:postitive_azimuth}(a) presents numerical results for $g_{\ua\da}/g= -0.2$. 
We find that the functions $\{f_{i}^\mathrm{P/AP}(\theta)\}$ stay constant to a good accuracy 
consistent with the analytical expressions \eqref{eq: triangular f_12}. 
The figure also shows the rescaled functions [defined by the left- and right-hand sides of Eq.\ \eqref{eq:P-AP relation}], 
clearly demonstrating the rescaling relations \eqref{eq:P-AP relation}. 
The dimensionless elastic constants $C$ and $D$ thus take the same values for the two types of fields 
and are plotted as functions of $g_{\ua\da}/g$ in Fig.~\ref{fig:elastic constants}(a). 
Both constants are linear functions of $g_{\ua\da}/g$, 
which is consistent with the fact that the elastic energy is a linear function of $g_{\ua\da}/g$ 
for a fixed vortex-lattice structure (see also Fig. 4 of Ref.\ \cite{Oktel06}).
Thus the numerical results are consistent with the effective field theory in the case of overlapping triangular lattices.

\subsection{Interlaced lattices}\label{sec:aniso_others}

We have performed similar analyses for interlaced lattices as shown in Fig.\ \ref{fig:postitive_azimuth}(b)-(e). 
The functions $\{f_{i}^\mathrm{P/AP}(\theta)\}$ displayed in the figure show anisotropy 
except in the right panels for (b) interlaced triangular and (d) square lattices. 
These behaviors are consistent with the analytical results in Eqs.~\eqref{eq : parameter} and \eqref{eq:f-effective}. 
Indeed, we can fit the numerical data perfectly using Eq.~\eqref{eq:f-effective} 
if we determine $C_i^\mathrm{P/AP}~(i=1,2,3,4)$ and $D_i^\mathrm{P/AP}~(i=1,2,3)$ {\it separately} for parallel or antiparallel fields. 
Figure\ \ref{fig:elastic constants}(b)-(e) presents the determined constants $\{C_i\}$ and $\{D_i\}$. 
We note that the constant $C_4$, which is newly introduced in this work and originates from 
the coupling between the symmetric and antisymmetric vortex displacements $\uv_\pm$, 
is indeed nonvanishing for (b) interlaced triangular lattices. 

However, the rescaling relations \eqref{eq:P-AP relation} derived from the effective field theory 
do not hold in Fig.\ \ref{fig:postitive_azimuth}(b)-(e). 
It can also be seen in different values of the constants $\{C_i^\mathrm{P/AP}\}$ and $\{D_i^\mathrm{P/AP}\}$ 
between the parallel- and antiparallel-field cases in Fig.\ \ref{fig:elastic constants}(b)-(e). 
The difference between the two cases tends to increase with increasing the ratio $g_{\ua\da}/g>0$. 
Furthermore, the constants for (d) square lattices show nonlinear dependences on $g_{\ua\da}/g$, 
which is inconsistent with the expected linear dependences for a fixed vortex-lattice structure (see Fig. 6 of Ref.\ \cite{Oktel06}). 
These results cannot be explained within our effective field theory. 

As discussed in the last paragraph of Sec.\ \ref{sec:Effect_elastic}, 
the elastic constants should take the same values between the parallel- and antiparallel-field cases 
because of the exact correspondence of the GP energy functionals between the two cases \cite{Furukawa14}. 
Therefore, a possible insufficiency of our effective field theory would reside in 
how the elastic constants are related to the coefficients in the dispersion relations. 
We infer that the derivative expansions and the coarse graining of the variables done in the derivation of the effective Lagrangian 
should be improved for interlaced vortex lattices which have a finite displacement between the components. 

\section{Summary and outlook}\label{sec:summary}

We have studied collective excitation modes of vortex lattices in two-component BECs 
subject to synthetic magnetic fields in parallel or antiparallel directions. 
Our motivation for studying the two types of synthetic fields stems from 
the fact that they lead to the same mean-field ground-state phase diagram \cite{Furukawa14} consisting of 
a variety of vortex-lattice phases \cite{Mueller02,Kasamatsu03}---it is interesting to investigate 
what similarities and differences arise in collective modes. 
Our analyses are based on a microscopic calculation using the Bogoliubov theory 
and an analytical calculation using a low-energy effective field theory. 
We have found that there appear two distinct modes with linear and quadratic dispersion relations at low energies 
for all the lattice structures and for both types of synthetic fields. 
These dispersion relations show anisotropy that reflects the symmetry of each lattice structure. 
In particular, we have pointed out that the anisotropy of the quadratic dispersion relation for interlaced triangular lattices 
can be explained by the term in the elastic energy that mixes the symmetric and antisymmetric vortex 
displacements---such a term was missing in a previous study \cite{Oktel06}. 
We have also found that the low-energy spectra for the two types of synthetic fields are related by simple rescaling 
in the case of overlapping triangular lattices that appear for intercomponent attraction ($-1<g_{\ua\da}/g < 0$). 
However, contrary to the effective field theory prediction, 
such relations are found to break down for interlaced vortex lattices, 
which appear for intercomponent repulsion ($g_{\ua\da}/g > 0$) and involve a vortex displacement between the components. 
This indicates a nontrivial effect of an intercomponent vortex displacement on excitation properties 
that cannot be captured by the effective field theory developed in this paper. 
We have also found that the spectra exhibit unique structures of band touching at some high-symmetry points or along lines in the Brillouin zone. 
We have discussed their physical origins on the basis of fractional translation symmetries and the numerical data of the Bogoliubov Hamiltonian matrix.


The Bogoliubov excitation spectra studied in this work 
can be utilized to calculate the quantum correction to the ground-state energy due to zero-point fluctuations [see Eq.~\eqref{eq:H_gg}], 
where the correction is expected to be enhanced as the filling factor $\nu$ is reduced. 
Despite the exact equivalence of the mean-field ground states between the parallel- and antiparallel-field cases \cite{Furukawa14}, 
we have found quantitatively different Bogoliubov excitation spectra for the two cases as shown in Fig.~\ref{fig:spectrum}. 
It is thus interesting to investigate how quantum corrections affect the rich vortex-lattice phase diagrams in the two cases. 
The present work would be a step toward understanding how the systems evolve from equivalent phase diagrams in the mean-field regime 
to markedly different phase diagrams in the quantum Hall regime \cite{Furukawa14,Furukawa13,Regnault13,Geraedts17,Furukawa17} 
as the filling factor is lowered. 

\ack
The authors thank Kazuya Fujimoto and Daisuke A. Takahashi for stimulating discussions. 
This work was supported by 
KAKENHI Grant Nos.\ JP18H01145 and JP18K03446 and  
a Grant-in-Aid for Scientific Research on Innovative Areas ``Topological Materials Science'' (KAKENHI Grant No.\ JP15H05855) 
from the Japan Society for the Promotion of Science (JSPS), 
and the Matsuo Foundation. 
T.\ Y.\ and S.\ H.\ were supported by JSPS through the Program for Leading Graduate Schools (ALPS). 
S.\ H.\ also acknowledges support from JSPS fellowship (KAKENHI Grant No. JP16J03619).

\appendix

\section{Lowest-Landau-level magnetic Bloch states in terms of the Jacobi theta function}\label{app:mag_Bloch_theta}

Here we show that for $\Nvor\to\infty$, the LLL magnetic Bloch states \eqref{eq:mag_Bloch} discussed in Sec.~\ref{sec2 LLL} 
can be rewritten in a compact form using Jacobi's theta function. 
In the resulting expression \eqref{eq:mag_Bloch_theta}, we can see the equivalence of these states to  the vortex-lattice wave functions introduced by Mueller and Ho \cite{Mueller02}. 
Furthermore, the expression \eqref{eq:mag_Bloch_theta} is useful for plotting density profiles of the vortex lattices and the excitation modes 
as in Figs.\ \ref{fig:mode_lowE} and \ref{fig:mode_highsym}. 

To derive such a compact expression of Eq.\ \eqref{eq:mag_Bloch}, we first rewrite it as
\begin{equation}\label{eq:mag_Bloch_zeta}
 \sqrt{A} \Psi_{\kv\alpha} (\rv)  = e^{-\rv^2/4\ell^2}
 \left[\zeta(\kv)\right]^{-1/2}\zeta (-\kvt_\alpha) ,~~
 \kvt_\alpha =\kv-i\frac{\rv}{2\ell^2}-\frac{\epsilon_\alpha}{2\ell^2} \ev_z\times\rv.
\end{equation}
Next we rewrite the function $\zeta(\kv)$ defined in Eq.\ \eqref{eq:zeta_k} in terms of the theta function. 
To this end, we parametrize the primitive vectors of the vortex lattices as $\av_1=a(1,0)$ and $\av_2=a(\tau_1,\tau_2)$, 
and introduce the modular parameters $\tau=\tau_1+i\tau_2$ and $\taub=\tau_1-i\tau_2$; 
the area of the unit cell in Eq.~\eqref{eq:quan_flux} is then given by $a^2\tau_2$. 
In the limit $\Nvor\to\infty$, the function $\zeta(\kv)$ can be rewritten as
\begin{equation*}
 \begin{split}
\zeta(\kv)
&=\sum_\mv \exp \left[ -\frac{\pi}{2\tau_2}(m_1^2+|\tau|^2m_2^2+2\tau_1m_1m_2) - im_1\kv\cdot\av_1-im_2\kv\cdot\av_2 + i\pi m_1m_2\right]\\
&=\sum_{m_2} \exp \left( -\frac{\pi |\tau|^2}{2\tau_2} m_2^2 - im_2\kv\cdot\av_2\right) 
 \sum_{m_1} \exp \left[ -\frac{\pi}{2\tau_2} m_1^2 + im_1 \left( -\kv\cdot\av_1 + i\frac{\pi\taub}{\tau_2} m_2 \right)\right] \\
&=\sum_{m\in\Zbb} \exp \left( -\frac{\pi |\tau|^2}{2\tau_2} m^2 - im\kv\cdot\av_2\right) 
 \sqrt{2\tau_2} \sum_{n\in\Zbb} \exp \left[ -\frac{\tau_2}{2\pi} \left( -\kv\cdot\av_1+i \frac{\pi\taub}{\tau_2}m-2\pi n \right)^2 \right].
 \end{split}
\end{equation*}
In the last line, we have used 
\begin{equation*}
 \sum_{m\in\Zbb} e^{-\alpha m^2+i\beta m} = \sqrt{ \frac{\pi}{\alpha} } \sum_{n\in\Zbb} e^{-(\beta-2\pi n)^2/(4\alpha)}~(\alpha,\beta\in \mathbb{C},~\Re~ \alpha>0),
\end{equation*}
which is obtained by the Poisson resummation. 
Using Jacobi's theta function of the third type $\theta_3(w|\tau)=\sum_{m\in\Zbb} \exp\left(\pi i\tau m^2+ 2\pi iwm\right)$ and the relation 
$\theta_3(w+\tau n|\tau)
=\exp\left(-\pi i\tau n^2-2\pi iwn\right) \theta_3(w|\tau)~~(w\in\mathbb{C})$, 
we can further rewrite $\zeta(\kv)$ as
\begin{equation*}\label{eq:zeta_theta}
\begin{split}
\zeta(\kv)
&=\sqrt{2\tau_2} \exp\left[ -\frac{\tau_2}{2\pi}(\kv\cdot\av_1)^2 \right] 
\sum_{n\in\Zbb} \exp \left( -2\pi\tau_2 n^2-2\tau_2 n\kv\cdot\av_1\right)
 \theta_3\left(\frac1{2\pi} \kv\cdot(\taub\av_1-\av_2)+n\taub \bigg|-\taub\right)\\
&=\sqrt{2\tau_2} \exp\left[ -\frac{\tau_2}{2\pi}(\kv\cdot\av_1)^2 \right] 
 \theta_3\left(\frac1{2\pi} \kv\cdot(\tau \av_1-\av_2)\bigg|\tau\right)
 \theta_3\left(\frac1{2\pi} \kv\cdot(\taub\av_1-\av_2)\bigg|-\taub\right).
\end{split}
\end{equation*}
Using this and $\theta_3(w|\tau)=\theta_3(-w|\tau)~(w\in\mathbb{C})$
and introducing $z_\alpha=(x+i\epsilon_\alpha y)/a$, $\kappa_x=\tau_2k_xa/2\pi$, and $\kappa_\pm=\tau_2(k_x\pm ik_y)a/2\pi$,  
we can rewrite Eq.~\eqref{eq:mag_Bloch_zeta} as
\begin{equation}\label{eq:mag_Bloch_theta}
\begin{split}
 \sqrt{A} \Psi_{\kv\alpha}(\rv) 
 =& (2\tau_2)^{1/4}
 \exp\left[ \frac{\pi}{2\tau_2} (-|z_\alpha|^2+z_\alpha^2+4i\kappa_x z_\alpha-2\kappa_x^2) \right] 
 [\theta_3(i\kappa_+|\tau)\theta_3(i\kappa_-|-\taub)]^{-1/2}\\
 &\times\theta_3\left(\frac{1+\epsilon_\alpha}{2}z_\alpha+ i\kappa_+\biggl| \tau\right)
 \theta_3\left( \frac{1 -\epsilon_\alpha}{2}z_\alpha+i\kappa_- \biggl|-\taub\right).
\end{split}
\end{equation}
Although the entire expression looks involved, the spatial dependence is expressed 
in a more compact manner than the original expression \eqref{eq:mag_Bloch}. 
Specifically, for $\epsilon_\alpha=+1$, the spatial dependence occurs in the part 
$\exp\left[ \frac{\pi}{2\tau_2} (-|z_\alpha|^2+z_\alpha^2+4i\kappa_x z_\alpha-2\kappa_x^2) \right] \theta_3\left(z_\alpha+ i\kappa_+| \tau\right)$. 
From the property of the theta function, this expression is found to have periodic zeros at 
$z_\alpha=\left(n_1+\frac12\right)+\left(n_2+\frac12\right)\tau-i\kappa_+$ with $(n_1,n_2)\in\Zbb^2$, 
which is consistent with Eq. \eqref{eq:zeros_Bloch}.
If we set $\kv=-\frac{1}{2\ell^2}\ev_z\times(\av_1+\av_2)=\frac12(\bv_1-\bv_2)=\frac{\pi}{\pi_2a}(\tau_2,-1-\tau_1)$, this expression is rewritten as
\begin{equation}\label{eq:mag_Bloch_theta1}
\begin{split}
 &\exp\left[ \frac{\pi}{2\tau_2} \left(-|z_\alpha|^2+z_\alpha^2+2i\tau_2 z_\alpha-\frac{\tau_2^2}{2}\right) \right] 
 \theta_3\left(z_\alpha+ \frac{1+\tau}{2} \bigg| \tau\right)\\
 &=\exp\left[ \frac{\pi}{2\tau_2} \left(-|z_\alpha|^2+z_\alpha^2\right) +\frac{\pi i}{4} (2-\tau_1)\right] 
 \theta_1\left(z_\alpha | \tau\right),
\end{split}
\end{equation}
where we use Jacobi's theta function of the first type
\begin{equation*}
\begin{split}
\theta_1(w|\tau)
&=-i\sum_{m\in\Zbb+1/2}(-1)^{r-1/2}\exp\left(\pi i\tau r^2+2\pi wr \right)\\
&=\exp\left[\pi i\left( w+\frac{-2+\tau}{4} \right)\right]\theta_3\left(w+\frac{1+\tau}{2} \bigg| \tau\right)~(w\in\mathbb{C}).
\end{split}
\end{equation*}
Equation \eqref{eq:mag_Bloch_theta1} is equivalent to the vortex-lattice wave function 
of Mueller and Ho \cite{Mueller02} up to multiplication by a constant factor.

\section{Derivation of the interaction matrix element \eqref{eq:V_kkkk}}\label{app: calc}

Here we derive the representation \eqref{eq:V_kkkk} of the interaction matrix element from Eq.~\eqref{V alpha beta}. 
By rewriting the LLL magnetic Bloch state \eqref{eq:mag_Bloch} as
\begin{equation*}
 \begin{split}
 \Psi_{\kv\alpha} (\rv) 
 = \frac{1}{\sqrt{A\zeta(\kv)}}\sum_\mv (-1)^{m_1m_2} 
 \exp \left[ -\frac1{4\ell^2} (\rv^2+\rv_\mv^2) + \frac1{2\ell^2} \rv\cdot (\rv_\mv-i\epsilon_\alpha \rv_\mv\times\ev_z)+i\kv\cdot\rv_\mv \right],
 \end{split}
\end{equation*}
we can calculate the integral of the product of four wave functions in Eq.\ \eqref{V alpha beta} as
\begin{equation}\label{eq:IntMat_mj}
\begin{split}
 &\bigg[ \prod_j \zeta(\kv_j) \bigg]^{1/2} \int d^2\rv~ \Psi_{\kv_1\alpha}^*(\rv) \Psi_{\kv_2\beta}^*(\rv) \Psi_{\kv_3\beta} (\rv) \Psi_{\kv_4\alpha}(\rv)\\
 &= \frac{1}{A^2} \sum_{\{\mv_j\}} (-1)^{\sum_j m_{j1}m_{j2}} \int d^2\rv~ \exp\bigg[ -\frac1{\ell^2}\rv^2+\frac1{2\ell^2}\rv\cdot\sum_j (\rv_{\mv_j}-i\epsilon_j\rv_{\mv_j}\times\ev_z)\\
 &~~~~~~~~~~~~~~~~~~~~~~~~~~~~~~~~~~~~~~~~~~~~~
 -\frac{1}{4\ell^2}\sum_j\rv_{\mv_j}^2+i\sum_j\kvt_j\cdot\rv_{\mv_j} \bigg]\\
 &= \frac{1}{2A\Nvor} \sum_{\{\mv_j\}} (-1)^{\sum_j m_{j1}m_{j2}} 
 \exp\left[ F_{\alpha\beta} (\rv_{\mv_1},\rv_{\mv_2},\rv_{\mv_3},\rv_{\mv_4})  +i\sum_j \kvt_j\cdot\rv_{\mv_j} \right],
\end{split}
\end{equation}
where we define $(\epsilon_1,\epsilon_2,\epsilon_3,\epsilon_4):=(-\epsilon_\alpha,-\epsilon_\beta,\epsilon_\beta,\epsilon_\alpha)$, 
$\kvt_{1,2}:=-\kv_{1,2}$, $\kvt_{3,4}:=\kv_{3,4}$, and
\begin{equation*}
 F_{\alpha\beta} (\rv_{\mv_1},\rv_{\mv_2},\rv_{\mv_3},\rv_{\mv_4})
 :=\frac{1}{16\ell^2}\left[ \sum_j (\rv_{\mv_j}-i\epsilon_j\rv_{\mv_j}\times\ev_z) \right]^2 -\frac1{4\ell^2} \sum_j\rv_{\mv_j}^2.
\end{equation*}
Introducing $\nv_j=\mv_j-\mv_4~(j=1,2,3)$, $F_{\alpha\beta}(\rv_{\mv_1},\rv_{\mv_2},\rv_{\mv_3},\rv_{\mv_4})$ can be rewritten as
\begin{align*}
 &F_{\alpha\beta}(\rv_{\mv_1},\rv_{\mv_2},\rv_{\mv_3},\rv_{\mv_4})\\
 &=\frac{1}{16\ell^2}\left[4\rv_{\mv_4} +\sum_{j=1}^3 (\rv_{\nv_j}-i\epsilon_j\rv_{\nv_j}\times\ev_z) \right]^2 
   -\frac1{4\ell^2} \left[ \sum_{j=1}^3 (\rv_{\mv_4}+\rv_{\nv_j})^2+\rv_{\mv_4}^2 \right] \\
 &=-\frac{i}{2\ell^2} \rv_{\mv_4} \cdot \sum_{j=1}^3 \epsilon_j (\rv_{\nv_j}\times\ev_z) + \tilde{F}_{\alpha\beta}(\rv_{\nv_1},\rv_{\nv_2},\rv_{\nv_3})\\
 &=-i\pi \sum_{j=1}^3 \epsilon_j (m_{41}n_{j2}-m_{42}n_{j1}) + \tilde{F}_{\alpha\beta}(\rv_{\nv_1},\rv_{\nv_2},\rv_{\nv_3}),\\
\end{align*}
where we define 
\begin{equation*}
\begin{split}
 \tilde{F}_{\alpha\beta}(\rv_{\nv_1},\rv_{\nv_2},\rv_{\nv_3})
 &:= \frac{1}{16\ell^2}\left[ \sum_{j=1}^3 (\rv_{\nv_j}-i\epsilon_j\rv_{\nv_j}\times\ev_z) \right]^2
 -\frac1{4\ell^2} \sum_{j=1}^3 \rv_{\nv_j}^2\\
 &= \frac{1}{8\ell^2} \sum_{i<j} \left[ (1-\epsilon_i\epsilon_j)\rv_{\nv_i}\cdot\rv_{\nv_j} + i(\epsilon_i-\epsilon_j)(\rv_{\nv_i}\times\rv_{\nv_j})_z \right]
 -\frac1{4\ell^2} \sum_{j=1}^3 \rv_{\nv_j}^2.
\end{split}
\end{equation*}
Equation \eqref{eq:IntMat_mj} can then be rewritten as
\begin{equation*}
\begin{split}
 &\frac{1}{2A\Nvor} \sum_{\mv_4,\nv_1,\nv_2,\nv_3} (-1)^{m_{41}m_{42}+\sum_{j=1}^3 (m_{41}+n_{j1})(m_{42}+n_{j2})}
 (-1)^{-\sum_{j=1}^3 \epsilon_j (m_{41}n_{j2}-m_{42}n_{j1})}\\
 &~~~~~~\times \exp \left[ \tilde{F}_{\alpha\beta}(\rv_{\nv_1},\rv_{\nv_2},\rv_{\nv_3})+i\left(\sum_{j=1}^4\kvt_j \right)\cdot\rv_{\mv_4}+i\sum_{j=1}^3 \kvt_j\cdot\rv_{\nv_j}   \right]\\
 &=\frac{1}{2A}\delta_{\sum_{j=1}^4\kvt_j,\mathbf{0}}^\mathrm{P} \sum_{\nv_1,\nv_2,\nv_3} (-1)^{\sum_{j=1}^3 n_{j1}n_{j2}} 
 \exp\left[ \tilde{F}_{\alpha\beta}(\rv_{\nv_1},\rv_{\nv_2},\rv_{\nv_3}) +i\sum_{j=1}^3 \kvt_j\cdot\rv_{\nv_j} \right].
\end{split}
\end{equation*}
Therefore, the interaction matrix element can be expressed as in Eq.~\eqref{eq:V_kkkk} with
\begin{equation*}
\begin{split}
 & S_{\alpha\beta} (\kv_1,\kv_2,\kv_3)\\
 &=\sum_{\nv_1,\nv_2,\nv_3} (-1)^{\sum_j n_{j1}n_{j2}} \exp[\tilde{F}_{\alpha\beta}(\rv_{\nv_1},\rv_{\nv_2},\rv_{\nv_3})
  -i\kv_1\cdot\rv_{\nv_1}-i\kv_2\cdot\rv_{\nv_2}+i\kv_3\cdot\rv_{\nv_3}].
\end{split}
\end{equation*}

Let us focus on the case of parallel fields ($\epsilon_\ua=\epsilon_\da=+1$). 
In this case, the function $S_{\alpha\beta}(\kv_1,\kv_2,\kv_3)$ depends on neither $\alpha$ nor $\beta$, and therefore we drop the subscripts $\alpha,\beta$. 
Using
\begin{equation*}
 4\ell^2 \tilde{F}(\rv_{\nv_1},\rv_{\nv_2},\rv_{\nv_3}) 
 =-\sum_j \rv_{\nv_j}^2
 +(\rv_{\nv_2}\cdot\rv_{\nv_3}+\rv_{\nv_1}\cdot\rv_{\nv_3})  
 -i (\rv_{\nv_2}\times\rv_{\nv_3}+\rv_{\nv_1}\times\rv_{\nv_3})_z ,
\end{equation*}
we find
\begin{equation}\label{eq:S_n_zeta}
\begin{split}
 S (\kv_1,\kv_2,\kv_3)
=& \sum_{\nv} (-1)^{n_1n_2} \exp\left( -\rv_\nv^2/4\ell^2 +i\kv_3\cdot\rv_\nv \right) \\
 & \times \zeta \left(\kv_1+(\rv_\nv\times\ev_z+i\rv_\nv)/4\ell^2\right) \zeta \left(\kv_2+(\rv_\nv\times\ev_z+i\rv_\nv)/4\ell^2\right),
\end{split}
\end{equation}
where the sums over $\nv_1$ and $\nv_2$ are rewritten in terms of $\zeta(\kv)$ in Eq.~\eqref{eq:zeta_k}, 
and the remaining dummy variable $\nv_3$ is replaced by $\nv$. 
We can further rewrite this by exploiting the following property of $\zeta(\kv)$ for $\sv\in\Zbb^2$: 
\begin{equation*}
\begin{split}
 &\zeta \left(\kv +(\rv_\sv\times\ev_z+i\rv_\sv)/2\ell^2 \right)\\
 &=\sum_\mv (-1)^{m_1m_2} \exp \left[ (-\rv_\mv^2+2\rv_\mv\cdot\rv_\sv)/4\ell^2 -i(\rv_\mv\times\rv_\sv)_z/2\ell^2-i\kv\cdot\rv_\mv \right]\\
 &=\sum_\mv (-1)^{m_1m_2} \exp \left[ -(\rv_\mv-\rv_\sv)^2/4\ell^2+\rv_\sv^2/4\ell^2 -i\pi (m_1s_2-m_2s_1) -i\kv\cdot\rv_\mv \right]\\
 &=(-1)^{s_1s_2} \exp\left(\rv_\sv^2/4\ell^2-i\kv\cdot\rv_\sv\right) \sum_\mv (-1)^{(m_1-s_1)(m_2-s_2)} \exp \left[ -(\rv_\mv-\rv_\sv)^2/4\ell^2-i\kv\cdot (\rv_\mv-\rv_\sv) \right]\\
 &=(-1)^{s_1s_2} \exp\left(\rv_\sv^2/4\ell^2-i\kv\cdot\rv_\sv\right) \zeta(\kv).
\end{split}
\end{equation*}
By setting $\nv=2\sv+\pv$ with $\sv\in \Zbb^2$ and $\pv\in\{0,1\}^2$, Eq.~\eqref{eq:S_n_zeta} can be rewritten as
\begin{equation*}
\begin{split}
 &S(\kv_1,\kv_2,\kv_3)\\
 &=\sum_{\pv\in\{0,1\}^2} \sum_\sv (-1)^{p_1p_2} 
 \exp\Big[-(2\rv_\sv+\rv_\pv)^2/4\ell^2+i\kv_3\cdot(2\rv_\sv+\rv_\pv) +\rv_\sv^2/2\ell^2  \\
 &~~~~~~~~~~~-i(\kv_1+\kv_2)\cdot\rv_\sv -i(\rv_\pv\times\ev_z+i\rv_\pv)\cdot\rv_\sv/2\ell^2 \Big]\\
 &~~~~~~\times\zeta \left(\kv_1+(\rv_\pv\times\ev_z+i\rv_\pv)/4\ell^2\right) \zeta \left(\kv_2+(\rv_\pv\times\ev_z+i\rv_\pv)/4\ell^2\right)\\
 &=\sum_{\pv\in\{0,1\}^2} (-1)^{p_1p_2} \exp \left( -\rv_\pv^2/4\ell^2 + i\kv_3\cdot\rv_\pv \right) \tilde{\zeta} \left(\kv_1+\kv_2-2\kv_3+(\rv_\pv\times\ev_z-i\rv_\pv)/2\ell^2 \right)\\
 &~~~~~~\times\zeta \left(\kv_1+(\rv_\pv\times\ev_z+i\rv_\pv)/4\ell^2\right) \zeta \left(\kv_2+(\rv_\pv\times\ev_z+i\rv_\pv)/4\ell^2\right).
\end{split}
\end{equation*}

In the case of antiparallel fields, $S_{\ua\ua}(\kv_1,\kv_2,\kv_3)$ is given by $S(\kv_1,\kv_2,\kv_3)$ shown above. 
The other $S_{\alpha\beta}(\kv_1,\kv_2,\kv_3)$'s can be obtained by using the relation $\Psi_{\kv\da}(\rv)=\Psi_{-\kv\ua}^\ast(\rv)$, 
leading to the result in Eq.~\eqref{eq:S_kkk^AP}.


\section{Fractional translation operators}\label{app:fractrans}

Here we give precise definitions of the fractional translation operators, 
$\Tcal^\para$ and $\Tcal^\apara$, which are introduced for the parallel- and antiparallel-field cases, respectively, in Sec.\ \ref{sec2 results}. 
We are concerned with the cases of (c) rhombic, (d) square, and (e) rectangular lattices. 
For these lattices, the wave vectors in Eq.\ \eqref{eq:qv_uada}, at which condensation occurs, are given by
$\qv_\ua = \epsilon_\ua \qv$ and $\qv_\da = -\epsilon_\da \qv$,  
where $\qv:=\ev_z\times\av_3/(4\ell^2) = (-\bv_1+\bv_2)/4$. 

To introduce the fractional translation, let us first recall that its square is equal to the translation by $\av_3$. 
For a single particle, the latter is expressed as $T_\alpha(\av_1) T_\alpha(\av_2)$. 
It acts on the magnetic Bloch states [with the shifted momenta as in Eq.~\eqref{eq:bt_Vt}] as 
\begin{equation}
 T_\alpha(\av_1) T_\alpha(\av_2) \Psi_{\kv+\qv_\alpha,\alpha}(\rv) 
 = e^{-i(\kv+\qv_\alpha)\cdot \av_3} \Psi_{\kv+\qv_\alpha,\alpha}(\rv) = e^{-i\kv\cdot \av_3} \Psi_{\kv+\qv_\alpha,\alpha}(\rv).
\end{equation}
Notably, the shift $\qv_\alpha$ does not appear in the eigenvalue $e^{-i\kv\cdot \av_3}$ 
since it is perpendicular to $\av_3$. 
The translation operator $T_\alpha(\av_1) T_\alpha(\av_2)$ can be rewritten as
\begin{equation}
 T_\alpha (\av_1) T_\alpha (\av_2) 
 = e^{-[\Kv_\alpha\cdot\av_1,\Kv_\alpha\cdot\av_2]/(2\hbar^2)} e^{-i\Kv_\alpha\cdot\av_3/\hbar}
 = e^{i\epsilon_\alpha\pi} T_\alpha (\av_3) = \Tt_\alpha{}^2,
\end{equation}
where $\Tt_\alpha:=e^{i\epsilon_\alpha \pi/2} T_\alpha(\av_3/2)$. 
In the following, we use $\Tt_\alpha$ in expressing the fractional translation. 

\subsection{Case of parallel fields}

\newcommand{\alphab}{{\bar{\alpha}}}
\newcommand{\betab}{{\bar{\beta}}}

In the case of parallel fields ($\epsilon_\ua=\epsilon_\da=1$), we can drop the subscript $\alpha$ in $T_\alpha(\sv)$ and $\Tt_\alpha$. 
To express the fractional translation, it is useful to modify the basis slightly from the magnetic Bloch states introduced in Sec.~\ref{sec2 LLL}. 
For the spin-$\da$ component, we use the same magnetic Bloch states as discussed in Sec.~\ref{sec2 LLL}. 
For the spin-$\ua$ component, we define $\Psi_{\kv+\qv,\ua} (\rv)$ by operating $\Tt$ on $\Psi_{\kv-\qv,\da}(\rv)$ as
\begin{equation} \label{eq:Tt_Psida}
 \Tt \Psi_{\kv-\qv,\da}(\rv) = e^{-i\kv\cdot\av_3/2} \Psi_{\kv+\qv,\ua} (\rv) .
\end{equation}
Using $T(\av_j)\Tt=e^{-[\Kv\cdot\av_j,\Kv\cdot\av_3/2]/\hbar^2} \Tt ~T(\av_j)=-\Tt ~T(\av_j)~(j=1,2)$, 
one can confirm that $\Psi_{\kv+\qv,\ua} (\rv)$ defined in this way has the expected momentum:
\[
 T(\av_j) \Psi_{\kv+\qv,\ua}(\rv) 
 = - e^{-i(\kv-\qv)\cdot\av_j} \Psi_{\kv+\qv,\ua}(\rv) = e^{-i(\kv+\qv)\cdot\av_j} \Psi_{\kv+\qv,\ua}(\rv) .
\]
Furthermore, by operating $\Tt$ on $\Psi_{\kv+\qv,\ua} (\rv)$, we have
\begin{equation} \label{eq:Tt_Psiua}
 \Tt \Psi_{\kv+\qv,\ua} (\rv) 
 = e^{i\kv\cdot\av_3/2} \Tt^2 \Psi_{\kv-\qv,\da}(\rv)  
 = e^{-i\kv\cdot\av_3/2} \Psi_{\kv-\qv,\da}(\rv)  .
\end{equation}
Equations \eqref{eq:Tt_Psida} and \eqref{eq:Tt_Psiua} indicate that the operator $\Tt$ has the role of interchanging 
$\Psi_{\kv-\qv,\da}(\rv)$ and $\Psi_{\kv+\qv,\ua} (\rv)$ 
with the multiplication of the same phase factor $e^{-i\kv\cdot\av_3/2}$, which is a useful feature of the present basis. 
In this representation, one can show
\begin{equation}\label{eq:Vt_sprv}
\begin{split} 
 &\Vt_{\alpha\beta} (\kv_1,\kv_2,\kv_3,\kv_4) \\
 &= \int\!\! d\rv d\rv' \Psi_{\kv_1+\qv_\alpha,\alpha}^*(\rv) \Psi_{\kv_2+\qv_\beta,\beta}^*(\rv')  
 g_{\alpha\beta} \delta^{(2)}(\rv-\rv') 
 \Psi_{\kv_3+\qv_\beta,\beta}(\rv') \Psi_{\kv_4+\qv_\alpha,\alpha}(\rv)\\
 &= e^{-i(\kv_1+\kv_2-\kv_3-\kv_4)\cdot\av_3/2}
 \int\!\! d\rv d\rv' \left[\Tt \Psi_{\kv_1+\qv_\alphab,\alphab}(\rv)\right]^* \left[\Tt \Psi_{\kv_2+\qv_\betab,\betab}(\rv')\right]^*\\
 &~~~~~~~~~~~~~~~~~~~~~~~~~~~\times g_{\alpha\beta} \delta^{(2)}(\rv-\rv') 
 \left[\Tt \Psi_{\kv_3+\qv_\betab,\betab}(\rv')\right] \left[\Tt \Psi_{\kv_4+\qv_\alphab,\alphab}(\rv)\right]\\
 &=e^{-i(\kv_1+\kv_2-\kv_3-\kv_4)\cdot\av_3/2} \Vt_{\bar{\alpha}\bar{\beta}} (\kv_1,\kv_2,\kv_3,\kv_4), 
\end{split}
\end{equation}
where the bars on $\alpha$ and $\beta$ indicate the spin reversal $\ua\leftrightarrow\da$ 
and we use the invariance of the interaction $g_{\alpha\beta}\delta^{(2)}(\rv-\rv')$ 
under the translation and the spin reversal ($g_{\alpha\beta}=g_{\bar{\alpha}\bar{\beta}}$). 

For a single particle, we define the fractional translation as the wave function changes by $\Tt$ in Eqs.~\eqref{eq:Tt_Psida} and \eqref{eq:Tt_Psiua}
followed by the spin reversal $\sigma_x$. 
For many particles, the fractional translation operator $\Tcal^\para$ can be expressed in the second-quantized form as
\begin{equation}\label{eq:T_P_b}
 \Tcal^\para \left( \bt_{\kv\ua}^\dagger, \bt_{\kv\da}^\dagger, \bt_{-\kv,\ua}, \bt_{-\kv,\da} \right) \Tcal^{\para \dagger} 
 = e^{-i\kv\cdot\av_3/2} \left( \bt_{\kv\ua}^\dagger, \bt_{\kv\da}^\dagger, \bt_{-\kv,\ua}, \bt_{-\kv,\da} \right) 
 \begin{pmatrix} \sigma_x & 0 \\ 0 & \sigma_x \end{pmatrix}  .
\end{equation}
Using Eq.~\eqref{eq:Vt_sprv}, one can confirm that the Bogoliubov Hamiltonian \eqref{eq:H_bb} is invariant under $\Tcal^\para$. 
The ground state $\ket{\mathrm{GS}}$ is obtained as the vacuum annihilated by the Bogolon annihilation operators $\gamma_{\kv,j}~(j=1,2)$ in Eq.~\eqref{eq:b_gamma}. 
The single-particle excitations $\gamma_{\kv,j}^\dagger \ket{\mathrm{GS}}~(j=1,2)$ 
can be used for the Bloch states $\ket{w_\kv^\pm}$ in the argument of Sec.\ \ref{sec2 results}. 

\subsection{Case of antiparallel fields}

In the case of antiparallel fields ($\epsilon_\ua = -\epsilon_\da = 1$), 
we again modify the basis slightly from the magnetic Bloch states introduced in Sec.~\ref{sec2 LLL}. 
While we use the same magnetic Bloch states as in Sec.~\ref{sec2 LLL} for the spin-$\da$ component, 
we define $\Psi_{\kv+\qv,\ua} (\rv)$ for the spin-$\ua$ component via
\begin{equation} \label{eq:Tt_Psidast}
 \Tt_\ua \Psi_{-\kv+\qv,\da}^*(\rv) = e^{-i\kv\cdot\av_3/2} \Psi_{\kv+\qv,\ua} (\rv) .
\end{equation}
Using $T_\alpha(\av_j)\Tt_\alpha=-\Tt_\alpha T_\alpha(\av_j)$ and $T_\alpha^*(\av_j)=T_{\bar{\alpha}}(\av_j)~(j=1,2;\alpha=\ua,\da)$, 
one can confirm that $\Psi_{\kv+\qv,\ua} (\rv)$ defined in this way has the expected momentum:
\[
 T_\ua (\av_j) \Psi_{\kv+\qv,\ua} (\rv) 
 = - e^{i\kv\cdot\av_3/2} \Tt_\ua [T_\da (\av_j) \Psi_{-\kv+\qv,\da}(\rv)]^*
 = e^{-i(\kv+\qv)\cdot\av_j} \Psi_{\kv+\qv,\ua} (\rv) .
\]
We also find
\begin{equation}\label{eq:Tt_Psiuast}
 \Tt_\da \Psi_{-\kv+\qv,\ua}^*(\rv) 
 = \Tt_\da \left[ e^{-i\kv\cdot\av_3/2} \Tt_\ua \Psi_{\kv+\qv,\da}^*(\rv) \right]^*
 = e^{-i\kv\cdot\av_3/2} \Psi_{\kv+\qv,\da} (\rv) .
\end{equation}
In this representation, one can show
\begin{equation} \label{eq:Vt_sprvcc}
\begin{split}
 &\Vt_{\alpha\beta} (\kv_1,\kv_2,\kv_3,\kv_4) \\
 &= e^{-i(\kv_1+\kv_2-\kv_3-\kv_4)\cdot\av_3/2}
 \int\!\! d\rv d\rv' \left[\Tt_\alpha \Psi_{-\kv_1+\qv,\alphab}^*(\rv)\right]^* \left[\Tt_\beta \Psi_{-\kv_2+\qv,\betab}^*(\rv')\right]^*\\
 &~~~~~~~~~~~~~~~~~~~~~~~~~~~\times g_{\alpha\beta} \delta^{(2)}(\rv-\rv') 
 \left[\Tt_\beta \Psi_{-\kv_3+\qv,\betab}^*(\rv')\right] \left[\Tt_\alpha \Psi_{-\kv_4+\qv,\alphab}^*(\rv)\right]\\
 &=e^{-i(\kv_1+\kv_2-\kv_3-\kv_4)\cdot\av_3/2} \Vt_{\bar{\alpha}\bar{\beta}}^* (-\kv_1,-\kv_2,-\kv_3,-\kv_4). 
\end{split}
\end{equation}

We define the fractional translation as the time reversal followed by the translation by $\Tt$. 
Here, the time reversal involves the complex conjugation, the wave vector reversal $\kv\to -\kv$ (about $\qv$), and the spin reversal $\ua\leftrightarrow\da$. 
In the second-quantized form, the fractional translation operator $\Tcal^\apara$ for many particles is represented as
\begin{equation}\label{eq:T_AP_b}
 \Tcal^\apara \left( \bt_{-\kv,\ua}^\dagger, \bt_{-\kv,\da}^\dagger, \bt_{\kv\ua}, \bt_{\kv\da} \right) \Tcal^{\apara \dagger} 
 = e^{-i\kv\cdot\av_3/2} \left( \bt_{\kv\ua}^\dagger, \bt_{\kv\da}^\dagger, \bt_{-\kv,\ua}, \bt_{-\kv,\da} \right) 
 \begin{pmatrix} \sigma_x & 0 \\ 0 & \sigma_x \end{pmatrix}  .
\end{equation}
Since $\Tcal^\apara$ is antiunitary, we find
\[
\left( \Tcal^\apara \right)^2 \left( \bt_{\kv\ua}^\dagger, \bt_{\kv\da}^\dagger, \bt_{-\kv,\ua}, \bt_{-\kv,\da} \right) \left(\Tcal^{\apara \dagger} \right)^2
= e^{-i\kv\cdot\av_3} \left( \bt_{\kv\ua}^\dagger, \bt_{\kv\da}^\dagger, \bt_{-\kv,\ua}, \bt_{-\kv,\da} \right), 
\]
by which we can confirm that $\left( \Tcal^\apara \right)^2$ is indeed equal to the translation by $\av_3$. 
By using Eq.~\eqref{eq:Vt_sprvcc}, we can also confirm that the Bogoliubov Hamiltonian \eqref{eq:H_bb} is invariant under $\Tcal^\apara$. 

Finally, we note that in the above argument, 
we have used $\sigma_x$ rather than the more standard one $i\sigma_y$ for the spin part of the time reversal. 
If we define ${\cal \Tt}^\apara$ by replacing $\sigma_x$ by $i\sigma_y$ in Eq.\ \eqref{eq:T_AP_b}, 
the original Hamiltonian \eqref{2ndQ hamiltonian} in the LLL basis is invariant under ${\cal \Tt}^\apara$. 
However, the Bogoliubov Hamiltonian \eqref{eq:H_bb} obtained after the breaking of U(1)$\times$U(1) symmetry as in Eq.\ \eqref{eq:b_cond}
is not invariant under ${\cal \Tt}^\apara$ because of the presence of the terms 
$\bt_{\kv\alpha}^\dagger \bt_{-\kv,\alpha}^\dagger$ and $\bt_{-\kv,\alpha} \bt_{\kv\alpha}$. 
Namely, the mixing of a particle and a hole in the Bogoliubov theory is in conflict with time-reversal symmetry in the standard form 
(see Ref.~\cite{Xu12} for a different type of conflict between condensation and time-reversal symmetry). 

\section{Excitation modes at high-symmetry points}\label{app: element}

In Sec.\ \ref{sec2 results}, we have discussed the origins of point and line nodes 
in the Bogoliubov excitation spectra in Fig.\ \ref{fig:spectrum}(c), (d), and (e) from the viewpoint of fractional translational symmetries. 
In Fig.\ \ref{fig:spectrum}, we further notice the following interesting features of the spectra at high-symmetry points: 
(i) coincidence of the excitation energies between the two types of fields at the $M_1$ and $M_2$ points for (c) rhombic, (d) square, and (e) rectangular lattices, 
and (ii) the point node at the $K_1$ point for (a) overlapping and (b) interlaced triangular lattices in antiparallel fields. 
We have not succeeded in explaining these features from a symmetry viewpoint. 
Here, we instead discuss their origins on the basis of the numerical data of the Bogoliubov Hamiltonian matrix $\mathcal{M}(\kv)$ and the density profiles of the excitation modes. 

\begin{figure}
\centering\includegraphics[width=15cm]{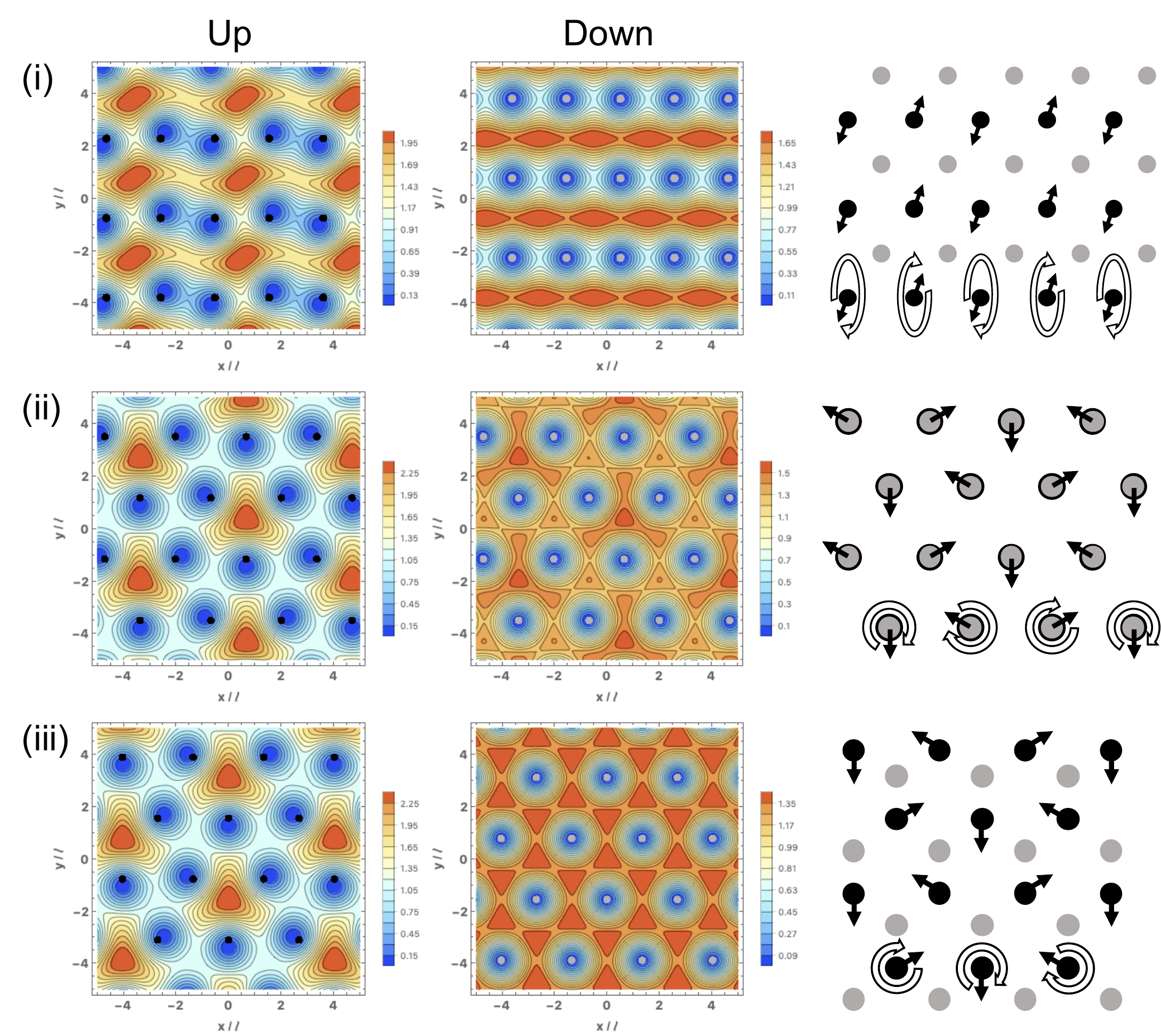} 
\caption{
Left and middle columns: density profiles $|\psi_\alpha(\rv,t=0)|^2/n$ ($\alpha=\ua,\da$) calculated using Eq.\ \eqref{eq:psi_rt_ki} with $c=0.3$ for the following three excitation modes: 
(i) the mode involving only the spin-$\ua$ component at the $M_1$ point for rectangular lattices, 
(ii) the mode involving a spin-$\ua$ particle and a spin-$\da$ hole at the $K_1$ point for overlapping triangular lattices in antiparallel fields, and 
(iii) the mode involving only a spin-$\ua$ particle at the $K_1$ point for  overlapping triangular lattices in antiparallel fields. 
In (i), the result is independent of the field direction for the spin-$\da$ component because of the decoupling of the two components.  
In (i) and (iii),  $|\psi_\da(\rv,t)|^2/n$ is the same as the ground-state density profile. 
Right column: schematic illustration of the vortex movement. 
Black (gray) circles indicate the locations of spin-$\ua$ ($\da$) vortices in the ground state (also shown in the other columns). 
Black arrows indicate the displacement of spin-$\ua$ vortices from the equilibrium positions at $t=0$, 
and empty arrows indicate their changes over the time interval $2\pi \hbar/E_i(\kv)$. 
We note that in (ii), spin-$\da$ vortices are also displaced in a way similar to spin-$\ua$ ones albeit with much smaller amplitudes. 
%
}\label{fig:mode_highsym}
\end{figure}

(i) The matrix $\mathcal{M} (\kv)$ at the $M_1$ point for (e) rectangular lattices is given by
\begin{equation}\label{eq:Mmat_rect_M1&M2}
\begin{split}
 \frac{2}{gn} \mathcal{M}(\kv)\bigg |_{M_1} =
  \begin{pmatrix}
  1.63 	     & 0 		   	 & 0.605-1.05i & 0 \\
  0 		     & 1.63 		    	 & 0 		       & 0.605 \mp 1.05i  \\
  0.605+1.05i & 0 		    	 & 1.63 	       & 0 \\
  0 		     & 0.605 \pm 1.05i & 0   	       & 1.63 \\
  \end{pmatrix} , \\
\end{split}
\end{equation}
where the upper and lower of the double signs correspond to the parallel- and antiparallel-field cases, respectively, 
and ``0'' indicates elements whose numerical values vanish with high accuracy. 
The structure of the matrix indicates that the spin-$\ua$ and $\da$ components are completely decoupled at this wave vector. 
We can thus construct the excitation mode involving only the spin-$\ua$ component, 
which is given by the vector $\left(\Ucal_{\ua },\Vcal_{\ua }\right)=(1.12,-0.248-0.431 i)$. 
For this mode, we present the density profiles $|\psi_\alpha(\rv,t=0)|^2/n~(\alpha=\ua,\da)$ and the schematic illustration of the vortex movement in Fig.\ \ref{fig:mode_highsym}(i). 
From this figure, we can interpret the decoupling of the two components in the following way: 
the forces acting on each spin-$\da$ vortex from the surrounding spin-$\ua$ vortices cancel out owing to the staggered nature of the displacement. 
Once the two components are decoupled in this way, they independently exhibit collective modes with identical spectra irrespective of the direction of the synthetic field.  
This explains the two-fold degeneracy of eigenenergies and the coincidence of those energies between the parallel- and antiparallel-field cases. 
Similar structures of the matrix $\mathcal{M} (\kv)$ are also seen at the $M_1$ point for (c) rhombic and (d) square lattices 
and at the $M_2$ point for (e) rectangular lattices. 

(ii) The matrix $\mathcal{M} (\kv)$ at the $K_1$ point for (a) overlapping triangular lattices in antiparallel fields  is given by
\begin{equation}\label{eq:Mmat_over_K1_AP}
\frac{2}{gn} \mathcal{M}(\kv)\bigg |_{K_1} =
  \begin{pmatrix}
   1.46   &  0         &  0	& -0.368 	\\
   0	     &  1.46   & -0.368 &  0  	\\
   0	     & -0.368 &  1.46 	&  0 		\\
  -0.368 &  0	   &  0&  1.46 	\\
  \end{pmatrix} .
\end{equation}
This matrix consists of two independent blocks---a block corresponding to a spin-$\ua$ particle and a spin-$\da$ hole 
and a block corresponding to a spin-$\da$ particle and a spin-$\ua$ hole. 
Since the two blocks have identical matrix elements, they show identical eigenenergies, which leads to the two-fold degeneracy at the $K_1$ point. 
For the mode involving a spin-$\ua$ particle and a spin-$\da$ hole [given by $\left(\Ucal_{\ua },\Vcal_{\da }\right)=(1.01, 0.129)$], 
we present the density profiles and the vortex movement in Fig.\ \ref{fig:mode_highsym}(ii), 
which exhibits a $\sqrt{3}\times\sqrt{3}$ structure reminiscent of the $120^\circ$ spin structure of an antiferromagnet on a triangular lattice. 
We note that the density changes and thus the amplitude of the vortex displacement are much smaller in the spin-$\da$ component 
than in the spin-$\ua$ component because $|\Vcal_\da|\ll |\Ucal_\ua|$. 

The matrix $\mathcal{M} (\kv)$ at the $K_1$ point for (b) interlaced triangular lattices in antiparallel fields is given by
\begin{equation}\label{eq:Mmat_int_tri_K1_AP}
\frac{2}{gn} \mathcal{M}(\kv)\bigg |_{K_1} =
  \begin{pmatrix}
  1.36 & 0      & 0	    & 0	\\
  0	  & 1.36 & 0 	    &  0  	\\
  0	  & 0      & 1.44   & 0.295 \\
  0      & 0      & 0.295 & 1.44	\\
  \end{pmatrix} .
\end{equation}
In this matrix, there is no coupling between a particle and a hole or between spin-$\ua$ and $\da$ particles. 
Thus, spin-$\ua$ and $\da$ particles exhibit independent excitation modes, which leads to the two-fold degeneracy at the $K_1$ point. 
For the mode involving only a spin-$\ua$ particle (given by $\Ucal_\ua=1$), 
we present the density profiles and the vortex movement in Fig.\ \ref{fig:mode_highsym}(iii); 
the spin-$\ua$ vortices are again found to exhibit a $\sqrt{3}\times\sqrt{3}$ structure. 
We note that in Eq.\ \eqref{eq:Mmat_int_tri_K1_AP}, there is a coupling between the spin-$\ua$ and $\da$ holes, 
which leads to excitations with non-degenerate negative eigenenergies; 
by performing the particle-hole transformation to these excitations, 
we obtain non-degenerate positive eigenenergies at the $K_2$ point, which is seen in Fig.~\ref{fig:spectrum}(b). 

Unfortunately, we have not been able to relate the vortex structures in Fig.\ \ref{fig:mode_highsym}(ii) and (iii) 
with the matrix structures in Eqs.~\eqref{eq:Mmat_over_K1_AP} and \eqref{eq:Mmat_int_tri_K1_AP}. 
At first sight, the cancellation of forces acting on a spin-down vortex from the surrounding spin-up vortices seem to occur in (iii); 
however, this assumption cannot explain why the block structure in Eq.~\eqref{eq:Mmat_int_tri_K1_AP} appears solely in the antiparallel-field case. 
Understanding the physical origins of the block structures in Eqs.~\eqref{eq:Mmat_over_K1_AP} and \eqref{eq:Mmat_int_tri_K1_AP} is thus still elusive. 

\section{Symmetry consideration of the elastic energy}\label{app:elastic}

\newcommand{\trm}{\mathrm{t}}
\newcommand{\Lambdat}{\tilde{\Lambda}}

Here we consider the elastic energy density $\Ecal_\elastic (\uv_\alpha,\partial_i\uv_\alpha)$ of the vortex lattices 
of two-component BECs shown in Fig.\ \ref{fig:vorlat}, 
and discuss how the symmetry constrains it into the form of Eqs.~\eqref{eq: eff energy elastic}, \eqref{eq : elastic energy}, and \eqref{eq : parameter}. 

We start from the quadratic forms of $\wv:=(w_1,w_2)^\trm$ and $\uv_-$: 
\begin{equation}
 \Ecal_\elastic^{(+)} = \frac{gn^2}{2} \wv^\trm C \wv,~~
 \Ecal_\elastic^{(-)}  = \frac{gn^2}{2\ell^2} \uv_-^\trm D \uv_-, ~~
 \Ecal_\elastic^{(+-)} = \frac{gn^2}{\ell} \wv^\trm F \uv_-,
\end{equation}
where $C$, $D$, and $F$ are real $2\times 2$ matrices, and $C$ and $D$ can be assumed to be symmetric. 
We assume that the vortex lattices have the symmetry under the coordinate transformation 
\begin{equation}
 \begin{pmatrix} x \\ y\end{pmatrix} \to 
 \begin{pmatrix} x' \\ y'\end{pmatrix} 
 =\Lambda \begin{pmatrix} x \\ y\end{pmatrix}. 
\end{equation}
Under this transformation, while $\uv_-$ is transformed by the same matrix $\Lambda$, 
$\wv$ is, in general, transformed by a different matrix $\Lambdat$. 
In order for the elastic energy to be invariant under this transformation, 
the following equations must be satisfied:
\begin{equation}\label{eq:CDF_inv}
 \Lambdat^\trm C \Lambdat = C,~~ \Lambda^\trm D \Lambda = D,~~ \Lambdat^\trm F \Lambda = F. 
\end{equation}

Here we consider the following transformations: 
\begin{align*}
 &\text{Rotation through the angle } \phi:~\Lambda=R(\phi)
 =\begin{pmatrix} \cos\phi & -\sin\phi \\ \sin\phi & \cos\phi \end{pmatrix},~
 \Lambdat=R(2\phi);\\
 &\text{Mirror about the $yz$ plane}:~\Lambda=M_x=\begin{pmatrix} -1 & 0 \\ 0 & 1 \end{pmatrix},~
 \Lambdat=\begin{pmatrix} 1 & 0 \\ 0 & -1 \end{pmatrix};\\
 &\text{Mirror about the $xz$ plane}:~\Lambda=M_y=\begin{pmatrix} 1 & 0 \\ 0 & -1 \end{pmatrix}, ~
 \Lambdat=\begin{pmatrix} 1 & 0 \\ 0 & -1 \end{pmatrix}.
\end{align*}
Each lattice structure in Fig.~\ref{fig:vorlat}(a)-(e) is invariant under the following transformation: 
\begin{equation*}
\text{(a)}~ R(\pi/3),M_x~~
\text{(b)}~ R(2\pi/3),M_x~~
\text{(c)}~ R(\pi)~~
\text{(d)}~ R(\pi/2),M_x,M_y~~
\text{(e)}~ R(\pi),M_x,M_y.
\end{equation*}
Requiring Eq.\ \eqref{eq:CDF_inv} for these transformations, we obtain a number of constraints on $C$, $D$, and $F$. 
For example, (i) the invariance under rotation through $\phi=\pi$ [satisfied by all but (b)], 
for which $\Lambda=-I$ and $\Lambdat=I$ (identity), immediately leads to $F=0$. 
(ii) The invariance under rotation through $\phi$ leads to 
\begin{equation*}
 (C_{11}-C_{22})\sin(2\phi)=C_{12}\sin(2\phi) =(D_{11}-D_{22})\sin \phi=D_{12}\sin\phi=0, 
\end{equation*}
which gives $C_{11}=C_{22}$ and $C_{12}=0$ for $\phi\ne n\pi/2$ and $D_{11}=D_{22}$ and $D_{12}=0$ for $\phi\ne n\pi$~($n\in \mathbb{Z}$). 
(iii) The invariance under the mirror reflection about the $yz$ plane leads to $C_{12}=D_{12}=F_{11}=F_{22}=0$. 
(iv) The invariance under rotation through $\phi=2\pi/3$ leads to $F_{12}=F_{21}$. 
Setting 
\begin{equation*}
 (C_1,C_2,C_3,D_1,D_2,D_3,F_1):=(C_{11},C_{22},2C_{12},D_{11},D_{22},2D_{12},2F_{12}),
\end{equation*}
and we finally obtain Eqs.\ \eqref{eq : elastic energy} and \eqref{eq : parameter}. 

\section*{References}

\bibliographystyle{iopart-num}


\bibliography{reference}

\end{document}